\documentclass[floatfix,aps,prd,abstract,10pt, twocolumn]{revtex4}
\usepackage{amsgen,amsmath,amstext,amsbsy,amsopn,amssymb}
\usepackage{graphicx}
\usepackage[usenames]{color}

\newcommand{\beq}{\begin{equation}}
\newcommand{\eeq}{\end{equation}}
\newcommand{\beqa}{\begin{eqnarray}}
\newcommand{\eeqa}{\end{eqnarray}}

\newcommand{\lexp}{\mathop{\langle}}
\newcommand{\rexp}{\mathop{\rangle}}

\font\BF=cmmib10

\def\k{{\hbox{\BF k}}}

\def\q{{\hbox{\BF q}}}

\def\la{\mathrel{\mathpalette\fun <}}
\def\ga{\mathrel{\mathpalette\fun >}}
\def\fun#1#2{\lower3.6pt\vbox{\baselineskip0pt\lineskip.9pt
       \ialign{$\mathsurround=0pt#1\hfill##\hfil$\crcr#2\crcr\sim\crcr}}}

\def\Mpc{\, h^{-1} \, {\rm Mpc}}

\def\kvecMpc{\, h \, {\rm Mpc}^{-1}}

\begin{document}

\title{Large-Scale Structure in Brane-Induced Gravity \\
II. Numerical Simulations}
\author{Kwan Chuen Chan} \email{kcc274@nyu.edu}
\author{Rom\'an Scoccimarro}
\affiliation{Center for Cosmology and Particle Physics, Department of
 Physics, \\  New York University, NY 10003, New York, USA}

\begin{abstract}
We use $N$-body simulations to study the nonlinear structure formation in brane-induced gravity, developing a new method  that requires alternate use of Fast Fourier Transforms and relaxation. This enables us to compute the nonlinear matter power spectrum and bispectrum, the halo mass function, and the halo bias.  From the simulation results, we confirm the expectations based on analytic arguments that the Vainshtein mechanism does operate as anticipated, with the density power spectrum approaching that of standard gravity within a modified background evolution in the nonlinear regime. The transition is very broad and there is no well defined Vainshtein scale, but roughly this corresponds to $k_*\simeq 2 \kvecMpc$ at redshift $z=1$ and $k_*\simeq 1 \kvecMpc$ at $z=0$. We checked that while extrinsic curvature fluctuations go nonlinear, and the dynamics of the brane-bending mode $C$ receives important nonlinear corrections, this mode does get suppressed compared to density perturbations, effectively decoupling from the standard gravity sector. At the same time, there is no violation of the weak field limit for metric perturbations associated with $C$. We find good agreement between  our measurements and the predictions for the nonlinear power spectrum presented in paper~I, that rely on a renormalization of the linear spectrum due to nonlinearities in the modified gravity sector. A similar prediction for the mass function shows the right trends. Our simulations also confirm the induced change in the bispectrum configuration dependence predicted in paper~I.

\end{abstract}

\maketitle

\section{Introduction}

There is strong observational evidence from different techniques that the universe is currently undergoing acceleration. The most studied  explanation of cosmic acceleration is that the universe is filled with an additional stress-energy component, usually called dark energy. The simplest example is the  cosmological constant, while more general possibilities include scalar fields with suitable potentials, as in early universe inflation. Since the study of the cosmology relies crucially on Einstein's general relativity (GR), and we test GR stringently only up to the scale of the solar system, it is possible that deviations from GR occur at cosmological scales. The observed cosmic acceleration may well be due to the fact that GR does not hold at such scales. It is thus important to explore predictions for theories that modify gravity at large scales to explain cosmic acceleration. 

Among these modified gravity theories, brane-induced gravity in five dimensions, the so-called Dvali-Gabadadze-Porrati (DGP) model
\cite{DGP} is one of the most widely studied. In this model, gravity is intrinsically five-dimensional, and a standard four-dimensional Einstein term is induced by the presence of standard model particles in a four-dimensional brane. Thus, while at small scales gravity reduces to GR, at large scales gravitons can leak out to the extra-dimension and the gravitational force law becomes five-dimensional. Interestingly, although it was not put forward to explain acceleration, it was later found that this model exhibits self-acceleration in the late universe~\cite{Deffayet}. This makes this model very interesting phenomenologically as it may explain the observed cosmic acceleration without invoking dark energy. In this paper we focus on the DGP model, but it is expected that the techniques developed here will be applicable to other models of massive gravity. For a review of the DGP model, see e.g.~\cite{Lue,Gabadadze,Koyama}.

Distinguishing dark energy from modified gravity is thus one of the main questions in understanding the cause of cosmic acceleration. At the level of expansion history alone, it is very difficult to do so, as any modified gravity expansion history can be mimicked by a suitable equation of state for the dark energy. However, the growth of perturbations will typically be different~\cite{LueSS}, unless one tunes the properties of the dark energy to make it cluster (instead of being smooth) on subhorizon scales to mimic a modified gravity model (see e.g.~\cite{KunzSapone} for such an attempt in linear perturbation theory); however,  such fine tuning may be even more contrived at the nonlinear level. In a companion paper (\cite{paper1}, hereafter paper I), we discuss predictions at the nonlinear level based on perturbation theory (see also~\cite{KTH} for closely related work), here we extend the work in paper I by running cosmological $N$-body simulations.

Studies of structure formation using $N$-body simulations in modified gravity has been carried out in~\cite{Stabenau, Laszlo,Shirata}, but were  restricted to linearizing the Poisson equation, while the field equations of viable modified gravity theories are typically nonlinear, since it is through nonlinearities how the extra degrees of freedom that modify gravity at large scales get suppressed at small scales, recovering GR as required by local tests. In~\cite{Oyaizu1,Oyaizu2, Schmidt}, $N$-body simulations are carried out for the $f(R)$ models, which involve nonlinear couplings of the scalar field to matter analogous (though very different in detail) to the second-derivative self-interactions present in the DGP model. 

In this work, we concentrate on fully nonlinear $N$-body simulations of the DGP model, solving numerically the equations of motion derived in paper~I~\cite{paper1} (see also~\cite{KoyamaSilva}).  The growth of perturbations in the linear and nonlinear case  was first studied using spherical symmetry in~\cite{LueSS}. The linear perturbations were studied beyond spherical symmetry by~\cite{KoyamaMaartens}, where the behavior in the bulk was established. All these treatments use the quasistatic approximation, appropriate for scales below the Hubble radius. For larger wavelengths, numerical solutions of linear theory were derived in~\cite{SSH07,Cardoso}. 

One of the interests of the present work is to study in some detail the so-called Vainshtein mechanism~\cite{Vainshtein} in the cosmological context. In paper~I we studied it analytically, here we resort to $N$-body simulations for a solution of the fully nonlinear equations that describe it. The Vainshtein mechanism was originally found in massive gravity. In the linearized case, the graviton propagator in the limit that the graviton mass goes to zero differs from the massless graviton theory: this is the well known vDVZ discontinuity~\cite{vDVZ}. As conjectured by Vainshtein~\cite{Vainshtein}, however, the discontinuity should disappear in the exact non-perturbative solution when nonlinearities  are included. While there have been some doubts in literature about whether the Vainshtein mechanism works in massive gravity (see e.g.~\cite{DKP}), recent work confirms its validity and clarifies the subtleties involved in matching the expected small-scale and large-scale behaviors~\cite{Babichev}.

The interest in the framework of cosmic acceleration is that this mechanism is responsible for modified gravity theories becoming GR at small scales, and the transition scale is important to pin down quantitatively to know where we can best test modifications of gravity in observations. In the DGP model, the Vainshtein effect is naturally built in~\cite{Deffayet02}. For a compact object, within its Vainshtein radius $r_{*} =( r_{\rm c}^2 r_{\rm sch} )^{1/3}$, where $ r_{\rm sch} $ is the Schwarzschild radius and $r_{\rm c}$ is the crossover scale, the extra scalar degree of freedom is strongly coupled and heavily suppressed, so that massless gravity is recovered in the nonlinear regime~\cite{Deffayet02,Gruzinov,Porrati, Tanaka04,GaIg05}.  In the case of cosmology, the precise value of $r_*$ is not as easy to obtain since it depends on the details of the fluctuation spectrum, while analytic estimates exist~\cite{LueSS,KoyamaSilva,paper1} its accurate determination requires a numerical simulation to cross check the analytic estimates. 

Our aim here is not to make precise predictions for the DGP model, as it is challenged from observations~\cite{Song,Wang,Fang} (see however~\cite{SDSSSNIa} for recent results with different conclusions) and theoretical considerations~\cite{DGPGhosts}, but rather understand the nonlinearities in the modified gravity sector in this model, since there are very good reasons to expect that it is a representative example of more complicated theories of massive gravity that may have better theoretical and observational properties, e.g.~\cite{deRham1,deRham2,Nicolis}. In particular, it is expected that the method introduced here for solving the nonlinear field equations would be useful in such models.

The rest of the paper is organized as follows. We first recall the background and perturbation equations for the nonlinear DGP model in
Sec.~\ref{sec:DGPeq}. In Sec.~\ref {sec:PMProcedure}, we outline the general procedures for $N$-body simulations in modified gravity. The
FFT-relaxation method is first described in Sec.~\ref{sec:FFT-relax} and then generalized in Sec.~\ref{sec:negativedis}.  A brief description of the computer resources requirement for the FFT-relaxation is also given at Sec.~\ref{sec:ComputingReq}. We present the power spectrum, bispectrum, halo mass function and halo bias from our simulations in Sec.~\ref{sec:Pk}, \ref{sec:Bispectrum}, \ref{sec:MassFunction}, and \ref{sec:Bias} respectively.  We conclude in Sec.~\ref{sec:Concl}.  For more discussion on technical aspects, in particular the ellipticity of the field equation,  and the comparison of the FFT-relaxation method with the standard Gauss-Seidel-relaxation (GS-relaxation) method, see Appendix~\ref{sec:GS-relax}.

\section{Equations of Motion}
\label{sec:DGPeq}

The background Friedmann equation in the DGP model in the self-accelerated branch is~\cite{Deffayet} 
\begin{equation}
\label{eq:DGPFriedmann}
H^2 = \frac{H}{r_{\rm c}} + \frac{8 \pi G}{3} \rho_{\rm m}, 
\end{equation} 
where $\rho_{\rm m}$ is the energy density of matter, and $r_{\rm c}$ is a cross-over scale which characterizes the length scale at which gravity goes from 4D to 5D behavior. In the long time limit, when matter density is sufficiently diluted, the universe enters a de Sitter phase with $H=r_{\rm c}^{-1}$. To explain the cosmic acceleration, $r_{\rm c}$ has to be of the order of the present Hubble radius. In particular, one can write the  current Hubble rate $H_0$ in terms of the present matter density parameter $\Omega_{\rm m}^0$ as
\begin{equation}
r_{\rm c} H_0 = \frac{ 1 }{ 1 - \Omega_{\rm m}^0}.
\end{equation}

The evolution of scalar metric perturbations in the subhorizon limit can be reduced to a coupled system of equations involving the Newtonian potential $\phi$ and a metric potential $C$ that characterizes the extrinsic curvature of the brane in the subhorizon limit. The modified Poisson equation reads~\cite{paper1}
\begin{eqnarray}
\label{eq:FullDGP1}
\bar{\nabla}^2 \phi  -  \frac{1}{\eta} \sqrt{ - \bar{\nabla}^2  } \phi
& +   &   \frac{1}{2 \eta} \bar{\nabla}^2 C   +     \nonumber    \\   
   \frac{ 3 \eta^2 - 5 \eta + 1 }{2 \eta^2  (2 \eta -1)  }  \sqrt{ - \bar{\nabla}^2  } C   &   = & \frac{3}{2} \frac{\eta  -1  }{\eta} \delta ,  
\end{eqnarray}
where $\delta$ is the density contrast and $C$ obeys a fully nonlinear partial differential equation
\begin{eqnarray}
\label{eq:FullDGP2}
(\bar{\nabla}^2 C)^2   &  + &   \alpha  \bar{\nabla}^2 C   - (\bar{ \nabla}_{ij} C)^2 +    \frac{ 3 \beta (\eta -1) }{2 \eta
-1 }    \sqrt{ - \bar{\nabla}^2  } C              \nonumber  \\ 
     & = & \frac{ 3( \eta -1 ) } {\eta } ( 1
- \beta \bar{\nabla}^{-1} ) \delta,  
\end{eqnarray}
see paper~I for a full derivation. Here we have defined 
\beq
\eta \equiv r_{\rm c} H, 
\eeq
\beq
\beta \equiv \frac{2\eta^2-2\eta-1}{\eta(2\eta-1)},\ \ \ \ \ \ \alpha \equiv \frac{3 (2\eta^2  - 2 \eta +1 ) }{\eta ( 2 \eta -1)}, 
\eeq  
and the dimensionless operator 
\beq
\bar{\nabla} \equiv \frac{ \nabla  }{ a H },
\eeq
has large eigenvalues ($k/aH \gg 1$) in the subhorizon limit. The nonlocal operators $ \sqrt{ - \bar{\nabla}^2  } $ and $\bar{\nabla}^{-1} \equiv  1/\sqrt{ - \bar{\nabla}^2  } $ can be easily handled as they are simply $\bar{k}$ and $\bar{k}^{-1}$ in Fourier space. The major obstacle in solving Eq.~(\ref{eq:FullDGP2}) stems from the nonlinear terms $( \bar{\nabla}^2C )^2$ and $(\bar{ \nabla}_{ij} C)^2$.  One of the main aims of this paper is to present a convergent algorithm to solve Eqs.~(\ref{eq:FullDGP1},\ref{eq:FullDGP2}).

It will also be useful to linearize the field equations since we shall compare the nonlinear solution against the result of linearizing the field equations, but keeping the equations of motion for matter fully nonlinear. After dropping the nonlinear terms, we can express $\bar{\nabla }^2 C  $ in terms of the density contrast $\delta$. We can then plug in the expression for $\bar{\nabla }^2 C$ into Eq.~(\ref{eq:FullDGP1}), and expanding the denominator in series of $\bar{k}^{-1} $ up to first order, we get the linearized Poisson equation
\begin{equation}
\label{eq:LinearDGP}
\nabla^2 \phi = 4 \pi G_{\rm eff} a^2 \bar{\rho}_{\rm m}\,  \delta, 
\end{equation}
where $G_{\rm eff}$ is an effective scale and time dependent Newton's constant
\beq
\left( {G_{\rm eff}\over G} \right)^{-1}= \frac{2 \eta    ( 2 + q ) -3 }{2 \eta ( 2 + q ) -4 }   +    \left(\frac{3 a }{k r_c }\right) \left(
   \frac{ \eta(1 +q ) -1 }{ \eta(2 +q ) - 2 } \right)^2
   \label{Geff}
\eeq
and $q$ is the {\em acceleration} parameter
\begin{equation}
q \equiv \frac{\ddot{a}{a}}{ \dot{a}^2 } = \frac{ 2 - \eta } { 2 \eta -1 }.
\end{equation} 
We emphasize that we can rigorously absorb the effects of modified gravity into a scale and time dependent effective gravitational constant only if we drop the nonlinear terms. When nonlinear effects are included, the arguments of paper~I suggest that one can represent most (though not all) of the nonlinear effects by a renormalization of $G$, part of the purpose of this paper is to verify those predictions against fully nonlinear numerical simulations. The first term in Eq.~(\ref{Geff}) represents the change in $G$ due to the linearized dynamics of $C$ (which mediates a repulsive force), while the second term inducing scale dependence represents the change in the gravitational force-law as one approaches the cross-over scale $r_c$.  Both of these effects make $G_{\rm eff}$ less than $G$. When the Vainshtein effect sets in due to nonlinearities in the dynamics of $C$, $G_{\rm eff}/G$ is driven to unity and one recovers the standard Poisson equation
\begin{equation}
\label{eq:GRPoisson}
\nabla^2 \phi = 4 \pi G a^2 \bar{\rho}_{\rm m}  \delta,
\end{equation}
but the expansion history will still be modified. Therefore, the benchmark to study the Vainshtein effect is to check whether the growth of perturbations at small scales approaches that of standard gravity with a modified expansion history.

\section{$N$-body Simulations}
\label{sec:Method}
We shall first briefly review the general method of particle mesh (PM) $N$-body simulations in Sec.~\ref{sec:PMProcedure}. We then introduce and describe the FFT-relaxation method we use in this paper in sections~\ref{sec:FFT-relax}-\ref{sec:ComputingReq}. For a comparison between this method and the standard Gauss-Seidel relaxation method see Appendix~\ref{sec:GS-relax}.

\subsection{PM $N$-body Simulations}
\label{sec:PMProcedure}
Since the procedure for running modified gravity $N$-body simulations is largely similar to the standard GR one, which is well-known, see e.g.~\cite{HockneyEastwood,Padmanabhan,Kravtsov}, we will only briefly outline it here. We also state the value of the cosmological and internal parameters used in our code. 

Our simulations use $256^3$ particles and start at $z=49$ for 128 Mpc/h box size and $z=24$ for the larger boxes. The initial conditions were generated using second order Lagrangian perturbation theory \cite{2lpt} with the transfer function from CMBFAST \cite{CMBFAST}. Unless otherwise noted our simulations assume $\Omega_m=0.27$, $h=0.7$ and $\Omega_b=0.04$.  The normalization is set by the present linear rms fluctuation $\sigma_8$, and we use $\sigma_8=0.9$ for the case of standard gravity, with the initial amplitudes the same for the different runs (thus $\sigma_8=0.715$ at $z=0$ for DGP models).

The computational domain is a cube with periodic boundary conditions. Due to the limitation of computer resources, we typically use $512^3$ grid points in our final results unless otherwise specified. Thus the number of particles are $1/8$ of the number of grid points. Each time step is uniform in $\ln a $, where $a$ is the scale factor, of width 0.005. When starting from $z=24$, it takes 618 time steps to reach $z=0$. 

The method we use is the standard Particle-Mesh (PM) algorithm. In each step we interpolate the mass to the grid points from the particles in order to calculate the potential and hence the force at the grid point. The force is then interpolated back to the particle from the grid points
using the same interpolation scheme. The assignment function we use is the Cloud-In-Cell (CIC), which is widely used in PM simulations. In 1D, the window function is  
\begin{equation}
W(x) = \left\{
\begin{array}{rl}
1-  \frac{|x|}{d}  & \text{if } |x|<d \\
0                  & \text{otherwise}
\end{array} \right. ,
\end{equation}
where $d$ is the size of the grid. For 3D, the assignment function is $W(x)W(y)W(z)$. Thus the mass distribution at the grid points $\rho_{\rm grid}$ can be given in terms of the particle distribution $\rho_{\rm particle}$ as 
\begin{eqnarray} 
\label{eq:CIC}
\rho_{\rm grid} (x,y,z)    =  \int dx' dy' dz'  \times   \qquad \qquad  
   \nonumber  \\ 
W(x-x')W(y-y')W(z-z')  \rho_{\rm particle} (x',y',z'). 
\end{eqnarray}

From the density contrast at the grid points, one can calculate the potential by solving the field equation. In GR, the field equation is just the Poisson equation, which can be solved by a single Fast Fourier Transform (FFT). In the DGP model, the field equations are nonlinear, and we will develop an algorithm to solve them in the next section.  

From the Newtonian potential $\phi$, we can advance the particles using Newton's law
\begin{eqnarray}
 \frac{d \bold{x}} {dt} & = &  \bold{u} , \\ 
\frac{d \bold{u}} {dt} & = & - 2 H \bold{u} - \frac{1}{a^2} \bold{\nabla} \phi ,  
\end{eqnarray}
where $\bold{x}$ is the comoving distance. The second equation holds in modified gravity because in a metric theory the motion of non-relativistic particles is sensitive to the Newtonian potential $\phi$ as in GR through the geodesic equation (with the background spacetime being FRW due to homogeneity and isotropy).

One can calculate $\bold{\nabla} \phi $ from $\phi(\bold{x})$ in real space using finite difference implementations of the gradient. But we choose instead to compute  $\bold{\nabla} \phi$ in Fourier space. In Fourier space,  $\bold{\nabla} \phi$ is just $ i\bold{k} \phi $, and we Fourier transform $ i\bold{k} \phi $ back to real space to get the force at the grid points. Again, we have to interpolate the force at the mesh points to the particles. To avoid self-force, we use the same assignment scheme as before, thus the force interpolation is the same as 
Eq.~(\ref{eq:CIC}), except that the roles of grid points and particles are interchanged. Once the force has been evaluated, we evolve the particle positions and velocities using the second-order leapfrog method. 

\begin{figure}[!t]
\centering
\includegraphics[angle =0,width=\linewidth]{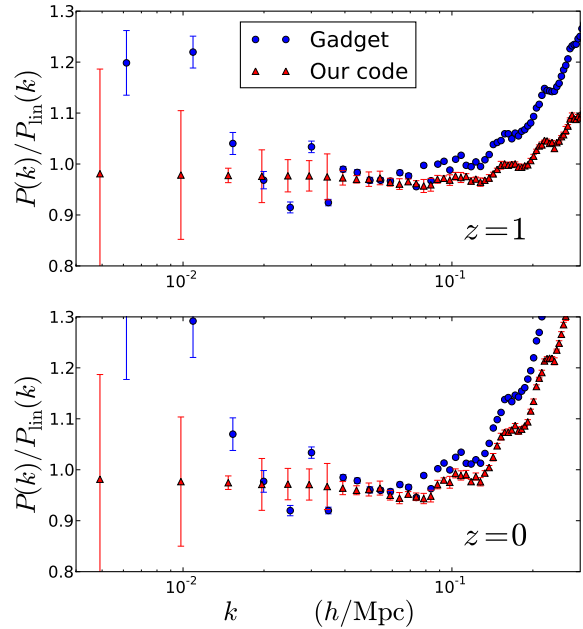}
\caption{ Comparison of the measured power spectrum from our code against the Gadget2 runs with the same initial conditions, but of different initial random seeds, for standard gravity. The upper panel is for $z=1$, and the lower panel corresponds to $z=0$. The Gadget power spectrum has been averaged over 49 realizations, while the spectrum from our code is from four realizations.  The numerical power spectra have been divided by the average initial spectrum used by our code scaled to the corresponding redshift using linear theory. }
\label{fig:GadgetVSGnspam}
\end{figure}

As a test, in Fig.~\ref{fig:GadgetVSGnspam}, we compare our code against Gadget2~\cite{Gadget} for the case of GR, where we show the ratio of the measured nonlinear power spectrum to the initial power spectrum scaled to the corresponding redshift by the linear growth factor $D_+(z)$ as a function of wavenumber $k$. In this test our code is run in a box of size $1280 \Mpc$ a side using $256^3$ particles. The power spectrum for Gadget has been averaged over 50 realizations using different initialization seeds (using $640^3$ particles in a $1280\Mpc$ box, see~\cite{RPTbao} for more details), and the associated error bars are one standard deviation on the mean of these realizations. The large-scale modulation of the Gadget result is mainly due to the fact that we divided by the average initial power spectrum used in our code, which corresponds to only four realizations.  The range in $k$ shown in Fig.~\ref{fig:GadgetVSGnspam} is chosen to go up to half  of the particle Nyquist frequency, within which the difference from Gadget is a systematically lower power by less than about 10\%.  Beyond the range shown in the figure, we find that our code gives significantly lower power than that from Gadget as expected from smoothing due to CIC interpolation. By taking the ratio of the power spectrum in Gadget to our code, we see that the ratio is constant with redshift, as expected from such smoothing. This systematic lower power have also been noted in many other PM simulations, e.g.~\cite{Stabenau,Laszlo,Oyaizu2} when making comparisons with high resolution codes or standard fitting formulae. In our case, these systematics are not too worrisome as we are interested in understanding differences (i.e. ratios) of power between GR, GR with modified expansion history (hereafter GRH), and modified gravity with linearized and full Poisson equation simulations, and for all of them we will use our code.

\subsection{FFT-Relaxation Method}
\label{sec:FFT-relax}
In this section, we describe the procedure that we use to solve Eqs.~(\ref{eq:FullDGP1}-\ref{eq:FullDGP2}). We shall first solve Eq.~(\ref{eq:FullDGP2}) to get $\bar{\nabla}^2 C $ as a function of $\delta$, and then substitute it into Eq.~(\ref{eq:FullDGP1}) to compute $\phi$. After obtaining $C$, Eq.~(\ref{eq:FullDGP1}) can be solved using standard FFT techniques.

To solve Eq.~(\ref{eq:FullDGP2}), the key idea is to notice that it can be regarded as a quadratic equation in $\bar{\nabla}^2 C $, which can
be solved for using the quadratic formula in terms of $(\bar{\nabla}_{ij}C)^2 $ and other remaining terms treated as sources. These sources, e.g. $\bar{\nabla}_{ij}C $ and $\sqrt{- \bar{\nabla}^2 C}$, depend in a non local way on  $\bar{\nabla}^2 C $. But these can be evaluated in Fourier space and an iterative relaxation method can be established among them.

Let us now discuss the details of the algorithm. We first start from the linearized version of Eq.~(\ref{eq:FullDGP2})
\begin{eqnarray}
\alpha \bar{\nabla}^2 C  &  + &  \frac{ 3 \beta (\eta -1) }{2 \eta
-1 }    \sqrt{ - \bar{\nabla}^2  } C  =         \nonumber   \\
&   &   \frac{ 3( \eta -1 ) } {\eta } ( 1 - \beta \bar{\nabla}^{-1} ) \delta.  
\end{eqnarray}
In Fourier space, this equation becomes 
\begin{eqnarray}
- \alpha \bar{k}^2 C +  \frac{ 3 \beta (\eta -1) }{2 \eta
-1 }   \bar{k}  C & = & \frac{ 3( \eta -1 ) } {\eta } ( 1
- \frac{\beta}{ \bar{k}} ) \delta.  
\end{eqnarray}
For notational convenience, we have used the same symbol for the quantity in real and Fourier space. From this Fourier representation, one can obtain $C(\bold{k})$ algebraically and this is used as our initial trial solution for the first time step in the  FFT-relaxation algorithm. Using that  
\begin{equation}
\nabla_{ij} C (\bold{k}) = - k_i k_j  C(\bold{k}), 
\end{equation}
and Fourier transforming $\nabla_{ij} C (\bold{k}) $ back to real space yields $\nabla_{ij} C (\bold{x}) $. From $\nabla_{ij} C (\bold{x})$ we
can then calculate  $( \nabla_{ij} C (\bold{x}) )^2 $, and then we can update $\nabla^2 C(\bold{x}) $ by treating Eq.~(\ref{eq:FullDGP2}) as a
quadratic equation in $\nabla^2 C(\bold{x}) $ using the quadratic formula
\begin{equation}
\label{eq:nabla^2C}
\nabla^2 C  = \frac{ - \alpha \pm \sqrt{\alpha^2 + 4 B } }{2},  
\end{equation}
with 
\begin{eqnarray}
\label{eq:B}
B \equiv ( \bar{\nabla}_{ij} C)^2  &  - &  \frac{ 3 \beta (\eta -1)
 }{2 \eta -1 } \sqrt{ - \bar{\nabla}^2  } C +   \nonumber   \\    
 \frac{ 3( \eta -1 ) } {\eta } ( 1  & - & \beta \bar{\nabla}^{-1} ) \delta.    
\end{eqnarray}
From Eq.~(\ref{eq:nabla^2C}), we see that the solution to the equation is not unique. However, the solution should reduce to that of the linearized equation in the linear limit and this holds only if we choose the plus sign. It is also not \textit{a priori} clear that the discriminant $\alpha^2 + 4 B$ is always positive. We will come back to this issue in Sec.~\ref{sec:negativedis}. 

With the new $\bar{\nabla}^2 C $, we can repeat the procedure to find $(\bar{\nabla}_{ij} C )^2$ and the nonlocal term, and therefore
$B$. Using the new $B$, we can compute $\bar{\nabla}^2 C $ again. This loop continues until some tolerance is met. That is, our relaxation algorithm can be written schematically as
\beq
\nabla^2 C^{(n)}  = \frac{ - \alpha + \sqrt{\alpha^2 + 4 B^{(n-1)} } }{2},  
\label{FFTrelax}
\eeq
at step $n$ in the iteration. Since each step involves calculating FFT's, we call our algorithm FFT-relaxation.

At each step we calculate the residual $R(\bold{x})$, which is the difference between the left hand side and right hand side of Eq.~(\ref{eq:FullDGP2}). A reference standard for comparing the size of the error is obtained by building the following quantity
\begin{equation}
\label{eq:normres}
\epsilon = \langle   \frac{ 3( \eta -1 ) } {\eta } |( 1- \beta \bar{\nabla}^{-1} ) \delta |  \rangle,  
\end{equation}
where the angular bracket denotes average over all the grid points. We then normalize the  average residual $\langle R \rangle $  to
$\epsilon$,  and monitor the \textit{normalized residual} $\langle R \rangle / \epsilon $ as the iteration progresses. We stop the iteration when the normalized residual is less than 1\%. After solving for $\bar{\nabla}^2 C$, we input it into Eq.~(\ref{eq:FullDGP1}) and solve for $\phi(\bold{k})$  in the Fourier space and then Fourier transform back $i\bold{k} \phi(\bold{k})$ to get the force. 

Except for the first time step in which we use the linear solution as the trial solution in the relaxation process, after the initial time step we use the previous time step nonlinear solution as the seed for the relaxation procedure. Using the previous nonlinear solution, it takes on average about three iterations for the normalized residual to decrease to 1\%.

\begin{figure}[!t]
\centering
\includegraphics[angle =0, width=\linewidth]{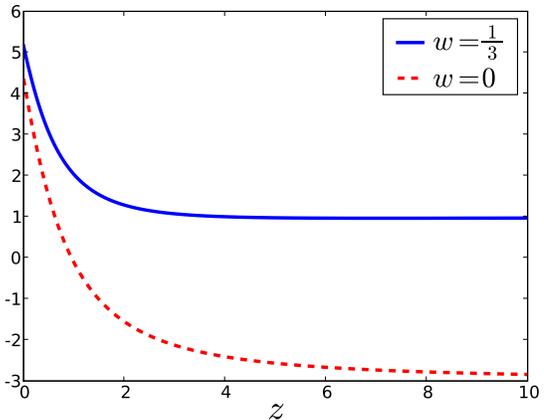}
\caption{Contributions to the discriminant $\Delta$ that do not depend on a trial solution for $C$, in voids ($\delta=-1$) as a function of $z$ for two choices of splitting the nonlinear terms. The top curve ($w=1/3$) corresponds to a splitting based on a multipolar expansion of the nonlinear kernel, while the bottom curve ($w=0$) denotes the naive splitting presented in Sec.~\protect\ref{sec:FFT-relax}. Note that in the latter case these contributions are negative already for $z>1$, and thus a good trial solution for $C$ is required to avoid $\Delta<0$.}
\label{fig:discriminant_check}
\end{figure}

\subsection{Splitting the Nonlinear Terms}
\label{sec:negativedis}
There is no guarantee that the discriminant $\Delta = \alpha^2 + 4 B  $ in Eq.~(\ref{eq:nabla^2C}) is always positive during time evolution, particularly in voids. To see more clearly the source of potential problems, let us rewrite Eq.~(\ref{eq:FullDGP2}) without the nonlocal terms, which we'll assume for simplicity here that do not alter things qualitatively (this is confirmed by numerical testing)
\beq
\label{eq:FullDGP2b}
(\bar{\nabla}^2 C)^2    +    \alpha \bar{\nabla}^2 C   - (\bar{ \nabla}_{ij} C)^2  = \frac{ 3( \eta -1 ) } {\eta }  \delta. 
\eeq
It is easy to see that the sum of the two quadratic terms  is positive definite (see Eq.~\ref{eq:NLFourier} below), therefore while in linear theory the left hand side of Eq.~(\ref{eq:FullDGP2b}) tracks the sign of $\delta$ (recall $\eta>1$) in the nonlinear case as voids empty and $\delta \to -1$ with $\nabla^2C$ becoming of order unity, the nonlinear term may take over and this equation may not have real solutions, depending on $\alpha(\eta)$. That would mean the theory is unphysical, or unstable in voids. However, even if this is not the case for the true solution, a particular implementation of the relaxation algorithm can be subject to these type of problems.

\begin{figure}[!htb]
\centering
\includegraphics[angle=0, width=\linewidth]{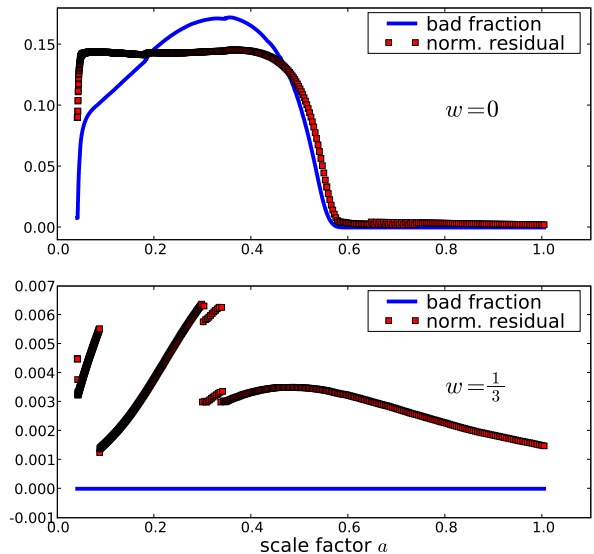}
\caption{The bad fraction and normalized residual for $w=0$ and 1/3 as a function of scale factor $a$. For $w=0$, the normalized residual is high when the bad fraction is non-zero in the range up to $a\sim0.6$, as expected from the arguments made in Sec.~\protect\ref{sec:FFT-relax}. The normalized residual can be easily reduced to below 1\% when the bad fraction is zero for the splitting choice of $w=1/3$. }
\label{fig:res_badfrac_1_0_1o3}
\end{figure}

To see this, let as rewrite the discriminant  $\Delta$ for Eq.~(\ref{eq:FullDGP2b})
\begin{equation}
\Delta =  \alpha^2 +4 \left[ ( \bar{\nabla}_{ij} C)^2  + \frac{ 3( \eta -1 )
 } {\eta } \delta \right]   .
 \label{eq:Discweq0}
\end{equation}
In Figure~\ref{fig:discriminant_check}, we plot the contributions to $\Delta$ that are independent of $C$, i.e. $ \alpha^2  +   12( \eta -1 )\eta^{-1 } \delta $ (which corresponds to $w=0$, with the parameter $w$ to be introduced shortly), with  $\delta = -1$ as a function of redshift $z$ for $\Omega_{\rm m}^0 =0.2$ (the precise value of $\Omega_{\rm m}^0 $ does not significantly change this). We see that for $z>1$, the value of   $ \alpha^2  + 12( \eta -1 )\eta^{-1 } \delta $ is less than 0 for $ \delta$ close to -1. This can potentially cause problems if the initial guess is not good enough and $ ( \bar{\nabla}_{ij} C)^2 $ is too small, making $\Delta $ negative. We shall see later that this heuristic argument is indeed quite accurate when we discuss Fig.~\ref{fig:res_badfrac_1_0_1o3}.

To make the sign of $\Delta$ less sensitive to our choice of the trial solution we split the nonlinear terms by introducing a parameter $w$ that separates how much of the local nonlinear piece $(\nabla^2C)^2$ is solved for immediately and how much is left for the iteration process. Specifically,  let us rewrite Eq.~(\ref{eq:FullDGP2}) as 

\begin{widetext}
\begin{eqnarray}
\label{eq:FullDGP2wpara}
(1-w)(  \bar{\nabla}^2 C )^2   +  \alpha  \bar{\nabla}^2 C  - \left[
  (\bar{ \nabla}_{ij} C)^2     -
 w(  \bar{\nabla}^2 C )^2    -  
 \frac{ 3 \beta (\eta -1) }{2 \eta
-1 }    \sqrt{ - \bar{\nabla}^2  } C  +  \frac{ 3( \eta -1 ) } {\eta } ( 1
- \beta \bar{\nabla}^{-1} ) \delta  \right]  =  0.   
\end{eqnarray}
\end{widetext}
With this simple splitting, our iterative procedure described in
Sec.~\ref{sec:FFT-relax} remains the same except Eqs.~(\ref{eq:nabla^2C}-\ref{eq:B}) now become  
\begin{equation}
\label{eq:nabla^2Cwpara}
\nabla^2 C  = \frac{ - \alpha + \sqrt{\alpha^2 + 4(1-w) B' } }{2 (1-w) },  
\end{equation}
and
\begin{eqnarray}
\label{eq:Bwpara}
B' &  \equiv &  ( \bar{\nabla}_{ij} C)^2   - w(  \bar{\nabla}^2 C )^2
 -  \frac{ 3 \beta (\eta -1) }{2 \eta -1 } \sqrt{ - \bar{\nabla}^2  }
 C     \nonumber       \\      
& + & \frac{ 3( \eta -1 ) } {\eta }
 ( 1 - \beta \bar{\nabla}^{-1} ) \delta.    
\end{eqnarray}
Neglecting the nonlocal term, the new discriminant reduces to 
\begin{equation}
\label{eq:NewDis}
\Delta =  \alpha^2 +4 (1- w)   \left[ ( \bar{\nabla}_{ij} C)^2   - w(
 \bar{\nabla}^2 C )^2  + \frac{ 3( \eta -1 )  } {\eta } \delta \right] .
\end{equation}
For $w$ in between 0 and 1, the factor $(1-w)$ reduces the magnitude of the term proportional to $\delta$, with the desirable effect that the contributions to $\Delta$ that do not depend on $C$ are now positive even for completely empty voids at all redshifts.  Indeed in Figure~\ref{fig:discriminant_check}, we also show $\alpha^2  +  8 ( \eta -1 )  \eta^{-1 } \delta $ (corresponding to $w=1/3$) with $\delta =-1$, and we see that this quantity is now positive for all $z$. Of course, now we have introduced a negative contribution into $\Delta$ given by $-w (\bar{\nabla}^2 C )^2 $ (which must be, since under correct choice of $C$ the result should be invariant under $w$) but now one expects the contribution coming from $C$ terms to be smaller in amplitude because of cancellations between the two terms that depend on $C$ and thus less sensitive to incorrect guesses coming from the trial solution.

How do we choose the arbitrary parameter $w$? The most physical choice we can think of  is  $w=1/3$, as it is motivated by the spherical collapse model. To see this we note that in Fourier space, the nonlinear terms in Eq.~(\ref{eq:FullDGP2}) can be written as
\begin{eqnarray}
\label{eq:NLFourier}
&  &   \Big[  (\bar{\nabla}^2C)^2    -   (\bar{\nabla}_{ij}C)^2 \Big](\bold{k})    =   \int   d^3k_1d^3k_2   \\  \nonumber  
&  &   \delta_{\rm D}(\bold{k}-\bold{k}_1-\bold{k}_2)\,   k_1^2 k_2^2\,  [1- \mu^2]\, C(\bold{k}_1)C(\bold{k}_2)\,   ,
\end{eqnarray} 
where $\mu \equiv \hat{\bold{k}}_1 \cdot \hat{\bold{k}}_2$.    We can decompose the angular dependence in multipoles and write $1-\mu^2 = (2/3) [1-{\cal P}_2(\mu)]$, where ${\cal P}_2(\mu)=(3\mu^2-1)/2$ is the second Legendre polynomial. Then,
\begin{equation}
\label{eq:MonQuad}
(\bar{\nabla}^2C)^2-(\bar{\nabla}_{ij}C)^2  = {2\over 3} (\bar{\nabla}^2C)^2 - {2\over 3} \hat{\cal P}_2\Big[\bar{\nabla}^2C,\bar{\nabla}^2C\Big],
\end{equation}
where we have introduced the operator $\hat{\cal P}_2$, such that
\begin{eqnarray}
\label{eq:P2def}
\hat{\cal P}_2[A,B]  &  \equiv  &   \int {d^3k \over (2\pi)^3}\, {\rm
 e}^{-i\bold{k}\cdot \bold{r}} \delta_{\rm
 D}(\bold{k}-\bold{k}_1-\bold{k}_2)\       \times  \\  \nonumber   
 &  &  {\cal P}_2(\mu)    \ A(\bold{k}_1) B(\bold{k}_2)\,   d^3k_1d^3k_2 
\end{eqnarray}
and in our case 
\begin{equation}
\label{eq:implementP2}
\hat{\cal P}_2\Big[\bar{\nabla}^2C,\bar{\nabla}^2C\Big]={3 \over 2}  (\bar{\nabla}_{ij}C)^2-{1\over 2} (\bar{\nabla}^2C)^2.
\end{equation}
Thus choosing $w=1/3$ means that we separate the nonlinear terms in a geometric way, and at each iteration we solve for the spherically averaged field equation treating the quadrupole as a source of perturbation. In other words, for $w=1/3$ the $C$ terms in Eq.~(\ref{eq:NewDis}) describe the quadrupole of the field $C$ which should be a small quantity (compared to the monopole) and thus making $\Delta$ more stable against choices of the trial solution for $C$. This decomposition is reminiscent to the approximations made in paper~I to initialize the resummation of the two-point propagator of the modified Poisson equation.

\begin{figure}[!htb]
\centering
\includegraphics[angle =0, width= \linewidth]{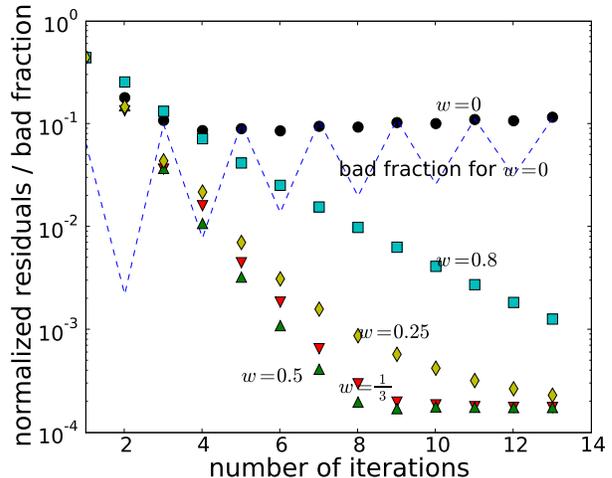}
\caption{Normalized residual of solving the nonlinear field equation for $w=0,~0.25,~1/3,~0.5,~0.8$ as a function of the number of iterations. For  $w=0$, we also show the variation of the bad fraction (dashed lines), while the bad fraction is always zero for all the other values of $w$. When bad fraction is non-zero ($w=0$), the normalized residual decreases for the first few iterations, and then it bounces back and increases gradually. The normalized residuals decreases most rapidly for $w=0.5$ and 1/3. The saturation at about $10^{-4}$ is probably due to the discretization of the equation, see text for discussion.}
\label{fig:w_testing}
\end{figure}

Another potentially interesting choice of $w$ is that which makes the discriminant $\Delta=0$ when $\delta = - 1$, while putting the $C$ terms to zero in Eq.~(\ref{eq:NewDis}). In this case $w$ is time dependent and given by 
\begin{equation}
\label{eq:ZeroDis}
w =  1 - \alpha^2 \frac{ \eta }{ 12 (\eta -1 ) }. 
\end{equation}
Figure~\ref{fig:w_testing} shows the normalized residual for different values of $w$. For the purpose of this illustration we have taken $z=24$ and a box size of 64 $\Mpc$ using a $256^3$ grid, although similar results hold for other choices (as long as $z>1$).  In all the the cases shown in Fig.~\ref{fig:w_testing}, everything is kept the same except the value of $w$. In particular, we start from the same linearized solution, and therefore the initial normalized residuals are the same independent of $w$.  

Let us denote those grid points at which the discriminant is negative as \textit{bad points}, and define the \textit{bad fraction} as the ratio of the number of bad points to the total number of grid points. For $w=0$, in addition to the normalized residual, we also show the bad fraction in dashed lines (the bad fraction is zero for all the other values of $w$ shown). When bad points appear, we set the discriminant in Eq.~(\ref{eq:nabla^2Cwpara}) artificially to zero so we can continue the iteration. For $w=0$ at the beginning the normalized residual (solid line) decreases as we iterate, but  it stops decreasing at some point, and increases gradually in a slightly oscillatory manner. The bad fraction (dashed line) also increases gradually but in a strongly oscillatory fashion. We found that the trend that the normalized residual decreases at first and then increases slowly is typical when the bad fraction is nonzero. 

We also plot $w=0.25$, which corresponds to using Eq.~(\ref{eq:ZeroDis}) for the particular $\eta$ for $z=24$, and it converges slightly slower than $w=0.5$ and the physically motivated $w=1/3$.  When $w$ is close to 1, such as $w=0.8$ shown here, the convergence is much slower, although the bad fraction is still zero. Although $w=0.5$ converges slightly faster than $w=1/3$, the difference is negligible in the range of the normalized residuals we are interested in, and $w=1/3$ is better motivated theoretically, thus we shall use $w=1/3$ in the subsequent computations.

The saturation of the normalized residual at about $1.5 \times 10^{-4}$ is most likely due to the discretization of Eq.~(\ref{eq:FullDGP2}). As the grid becomes more refined, the residual calculated using a \textit{consistent} numerical method should approach zero. However, this 
assumes that the fluctuation at the grid scale remains unchanged as the size of the grid increases. We find that the normalized residuals saturate to roughly the same value using $128^3$, $256^3$ and $512^3$ grids. However, as the size of the grid increases the fluctuations at the grid scale increase, since we probe smaller scales. When we increase the grid size and artifically decrease $\sigma_8$ to maintain the level of fluctuations at the grid scale constant, the saturated normalized residual value indeed decreases. We emphasize that in any case the saturated value seen in Fig.~\ref{fig:w_testing} is much lower than we require.

In Figure~\ref{fig:res_badfrac_1_0_1o3} we plot the bad fraction and normalized residual for $w =0$ (top panel) and $w=1/3$ (bottom panel) during a full simulation from $a=0.04$ to $a=1$ using $256^3$ grid points in a $64\Mpc$ box. Whenever the bad fraction is nonzero we terminate the iteration of the relaxation method when the minimum residual is reached. For $w=0$, the bad fraction is non-zero from the beginning $a=0.04$ to about $a=0.6$, and the minimum normalized residual that can be attained is also rather high. Note that the range where the bad fraction is nonzero agrees with the simple arguments leading to Eq.~(\ref{eq:Discweq0}) and displayed by the $w=0$ curve in Fig.~\ref{fig:discriminant_check}. When we set $w=1/3$, on the other hand, the bad fraction is always zero and the normalized residual can be maintained below 1\% during the whole simulation with only a few iterations of the relaxation procedure.

\subsection{Computational  Resources}
\label{sec:ComputingReq}

The computing time and memory required to perform this type of numerical simulations are much higher than the standard gravity $N$-body simulations. 
For example, because of non-local terms in Eq.~(\ref{eq:Bwpara}), we  Fourier transform $\delta $, $\nabla^2 C$, $(\nabla^2 C)^2$, and $(\nabla_{ij}C )^2$ to evaluate $B'$ in Fourier space. To compute $( \nabla_{ij}C(\bold{x} ) )^2 $ from $\nabla^2C(\bold{k})$ we have to loop over all $k$ modes and inverse FFT to real space six times. Including calculating residuals, we need to perform 15 FFTs in one iteration. For each time step in the simulation, we need on average three cycles to attain the desired accuracy. Thus we have to do about 45 FFTs to solve the field equations for $C$ in one time step, while in standard GR one FFT is required. Computing the force requires 3 more FFTs in either case. Therefore there is a factor of about $48/4=12$ in terms of cost from FFTs alone.

We have parallelized our code using OpenMP to speed up the computations. When the redshift is high, the nonlinearity is small, therefore we use the linearized DGP Poisson equation and switch to fully nonlinear calculations at some epoch depending on the size of the box, e.g. at $a=0.1$ for the $512 \Mpc$ box.  Using 4 CPUs, a simulation with $256^3$ particles in standard gravity takes about 10 hours, while for the nonlinear DGP model it takes about 4 days. When using a $512^3$ grid, the FFT-relaxation code requires about 16 GB of memory. Although our implementation is not very economical as far as computing resources is concerned, and can certainly be improved, it is more than enough for our purposes.

\section{The Power Spectrum}
\label{sec:Pk}

We now present results for the power spectrum of the density perturbations measured from simulations.  We would like to compare the power spectrum for three different models obtained by simulating different perturbation equations but with the same background evolution given by  the Friedmann equation Eq.~(\ref{eq:DGPFriedmann}), which we denote by 
\beq
P_{\rm nlDGP}(k), \ \ \ \ \ P_{\rm lDGP}(k), \ \ \ \ \ P_{\rm GRH}(k),
\eeq
corresponding to the fully nonlinear DGP model (Eqs.~\ref{eq:FullDGP1}-\ref{eq:FullDGP2}), the linearized DGP model (Eq.~\ref{eq:LinearDGP}), and GR (Eq.~\ref{eq:GRPoisson}) with the same $H(z)$, respectively. Since all three models share the same expansion history, the differences in their power spectra arise only from the different gravitational forces for perturbations. 

At large scales, we expect that $P_{\rm nlDGP}\simeq P_{\rm lDGP}$ as the linearized modified Poisson equation, Eq.~(\ref{eq:LinearDGP}), becomes a good approximation. On the other hand, at small scales we expect that $P_{\rm nlDGP}\simeq P_{\rm GRH}$ as the Vainshtein mechanism suppresses $C$ relative to $\phi$ and the standard Poisson equation becomes a good approximation. Therefore, the extra nonlinearities in the modified gravity sector mean that there are two models with the same linear spectrum but different nonlinear spectrum ($P_{\rm nlDGP}$ and $P_{\rm lDGP}$) and two models with different linear spectrum that have the same nonlinear spectrum ($P_{\rm nlDGP}$ and  $P_{\rm GRH}$).

\begin{figure*}[t!]
\begin{center}
\includegraphics[width=0.49\textwidth]{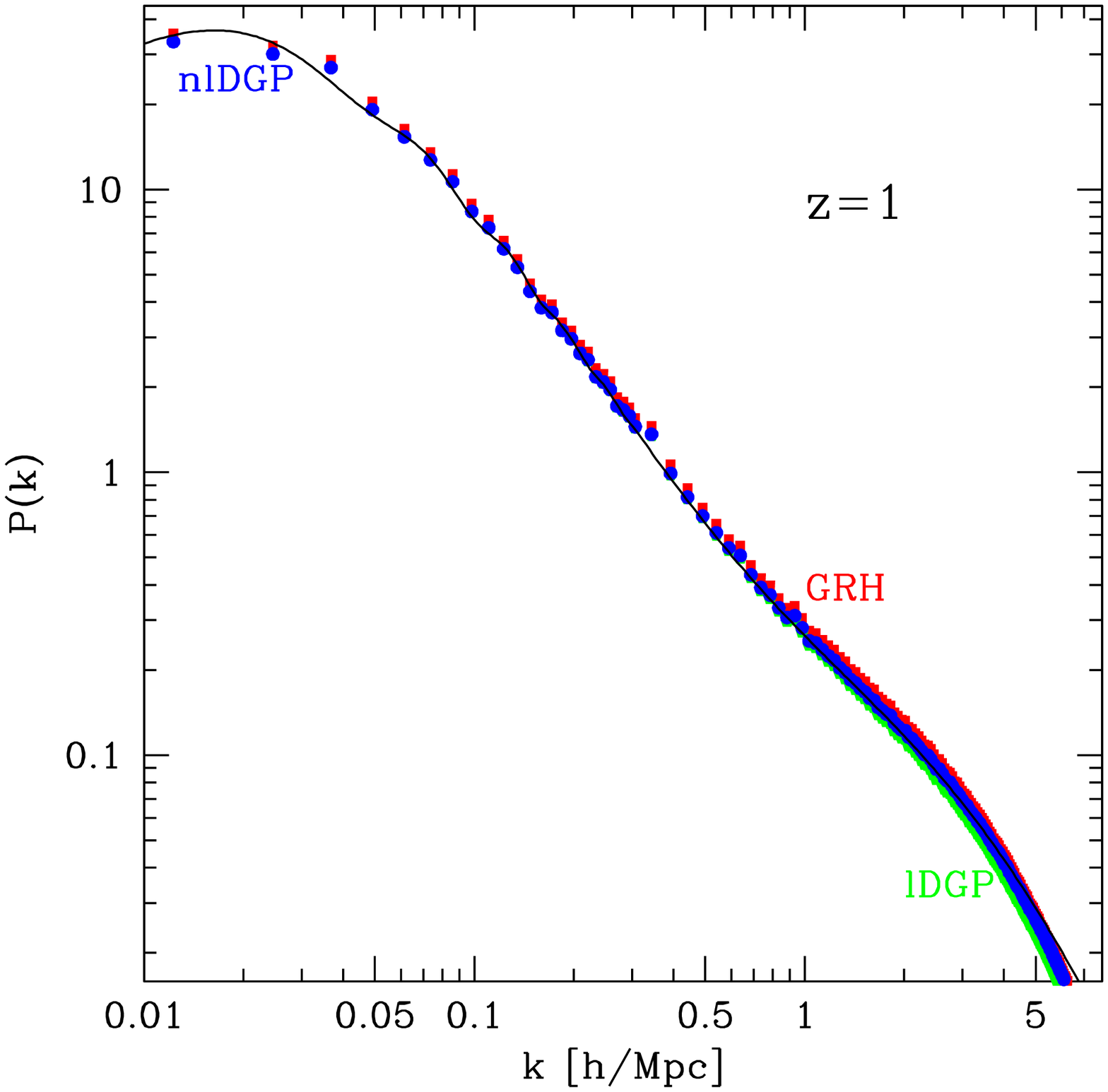}
\includegraphics[width=0.49\textwidth]{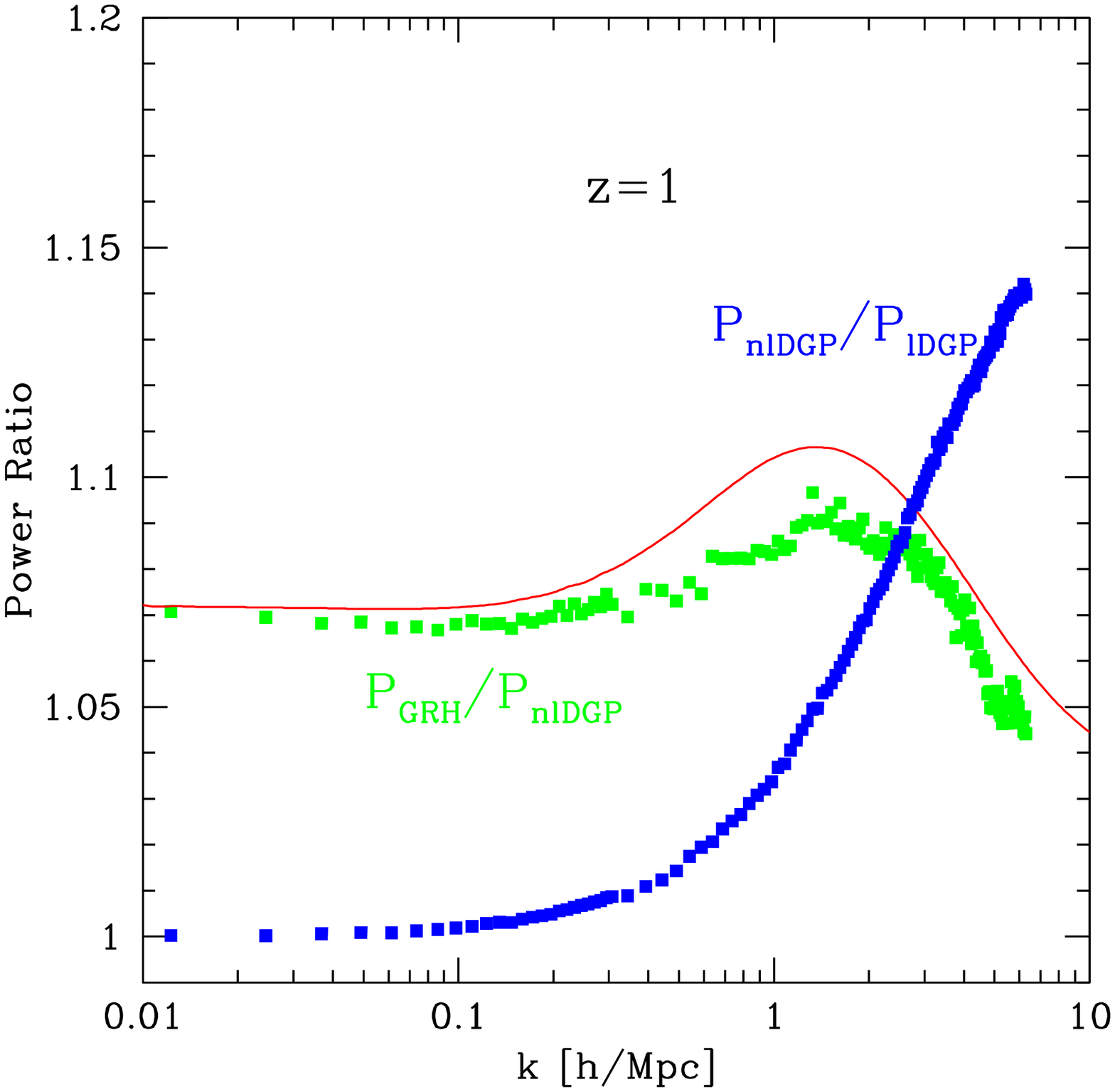} 
\caption{Dark matter power spectra from the nonlinear DGP model (nlDGP) , linear DGP (lDGP), and GR perturbations with the same expansion history (GRH) at $z=1$. The left panels show the power spectra, and the right panels shows ratios to better see the differences. Two sets of computational boxes are shown for each case, covering a different range in $k$ (see text). The solid line denotes the predictions from paper I for $P_{\rm nlDGP}$ (left panel) and $P_{\rm GRH}/P_{\rm nlDGP}$ (right panel).}
\label{fig:pk_z1}
\end{center}
\end{figure*}

\begin{figure*}[t!]
\begin{center}
\includegraphics[width=0.49\textwidth]{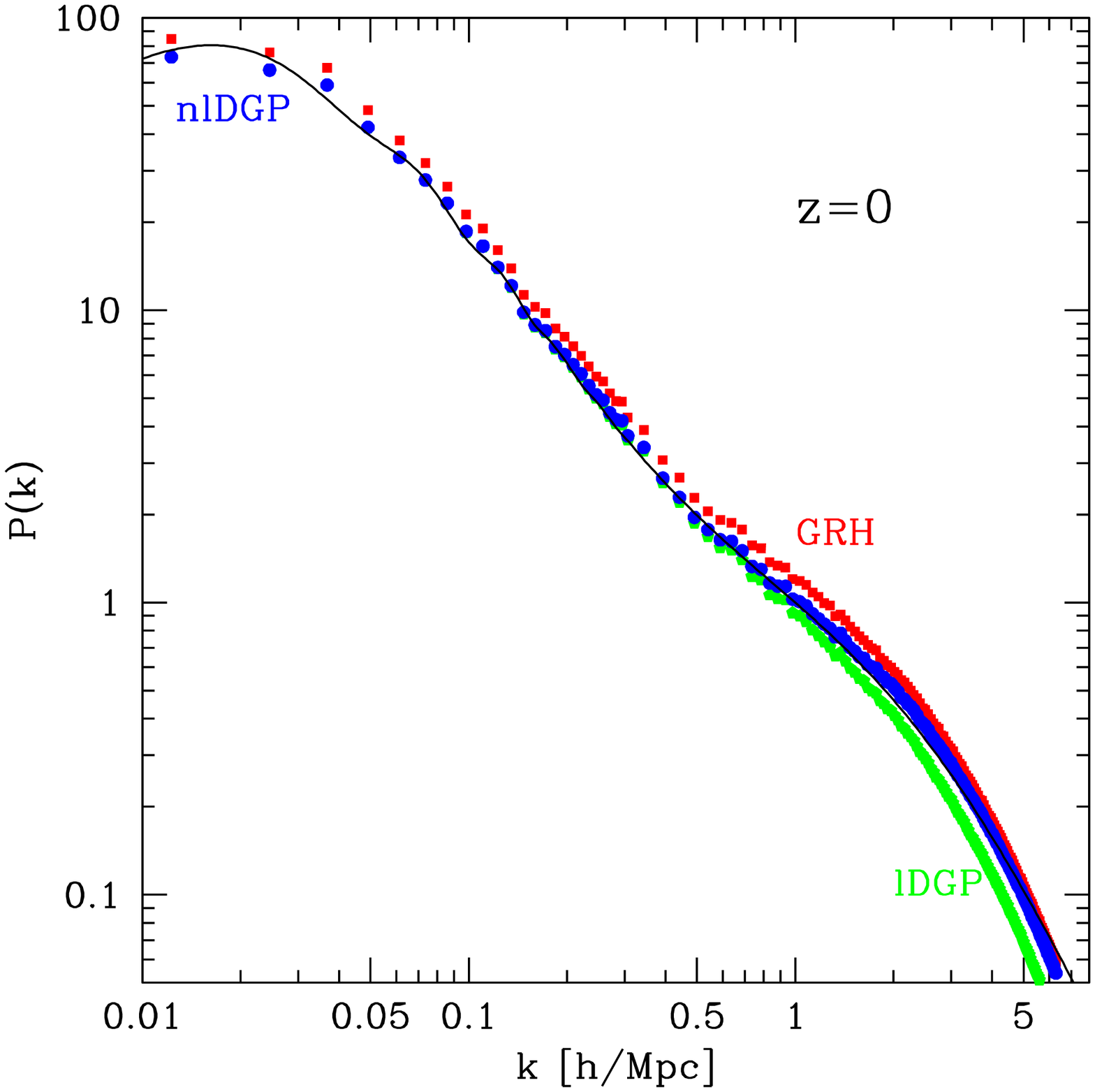}
\includegraphics[width=0.49\textwidth]{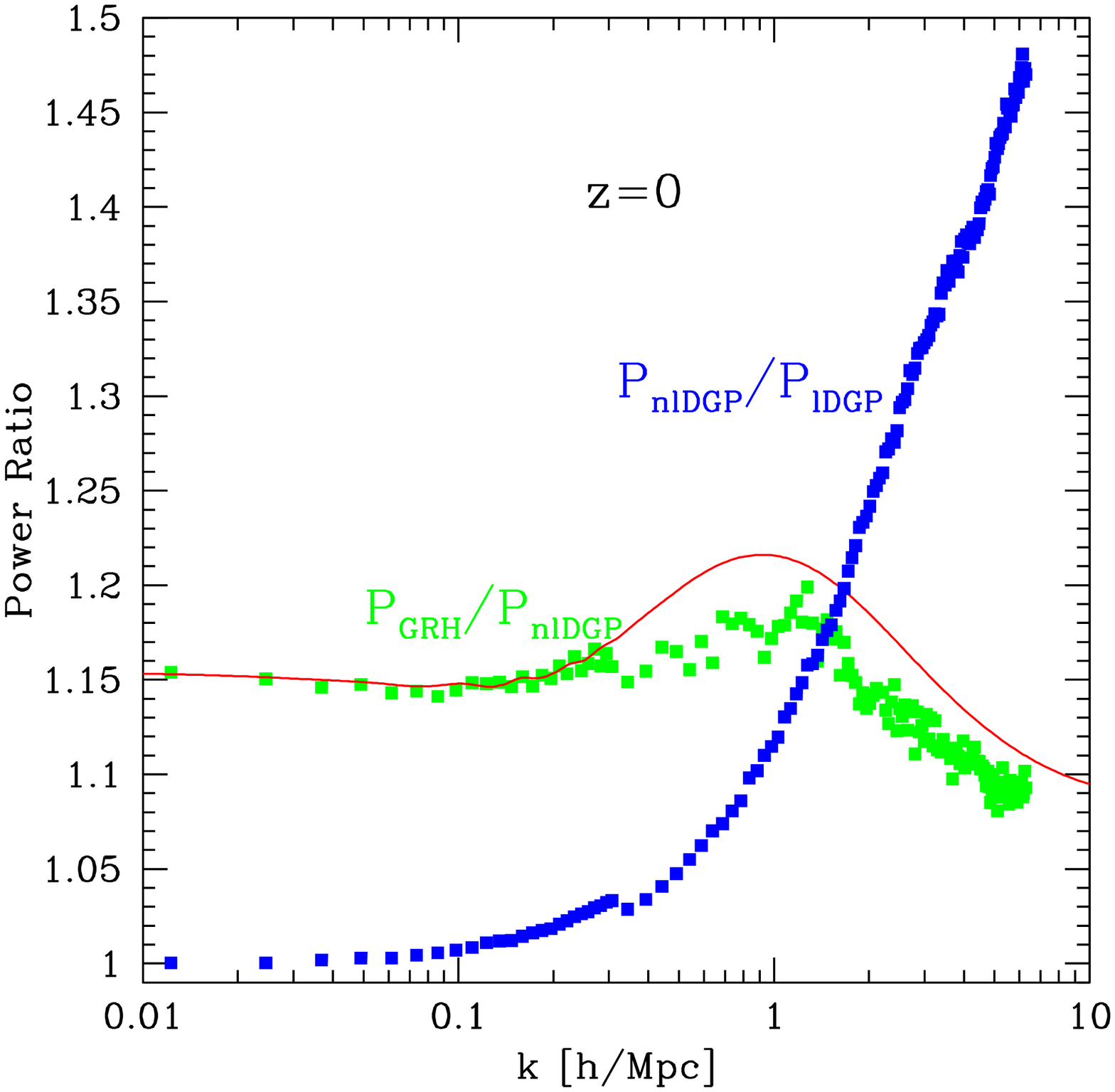} 
\caption{Same as Fig.~\ref{fig:pk_z1} but for $z=0$ }
\label{fig:pk_z0}
\end{center}
\end{figure*}

Figures~\ref{fig:pk_z1} and~\ref{fig:pk_z0} show the power spectra for these three models  at redshift $z=1,0$, respectively, with the left panels showing power spectra, and the right panels showing ratios to better see the details. Each set of symbols corresponds to merging two computational boxes, of size $512\Mpc$ (large scales up to $k\simeq 0.3\kvecMpc$) and $128\Mpc$ ($k\simeq 0.3\kvecMpc$ and up). While one should not take the absolute value of the power spectra too seriously at high frequencies (since suppression due to grid smoothing is significant) we can trust much more the relative differences between the spectra.  To see this better, the right panels show the ratios $P_{\rm nlDGP}/ P_{\rm lDGP}$ and $ P_{\rm GRH}/P_{\rm nlDGP} $. We see the expected behavior, $P_{\rm nlDGP}/ P_{\rm lDGP}$ becomes unity at large scales, and gets amplified at small scales as the Vainshtein effect makes gravity stronger. On the other hand, $P_{\rm GRH}/P_{\rm nlDGP} $ is larger than unity at large scales, as DGP gravity is weaker than GR, and the ratio is driven to unity at small scales once the Vainshtein mechanism sets in.  We see that the transition to GR is not complete at the smallest scales shown, even at $z=0$ and $k\simeq 6\kvecMpc$ there is still about 10\% enhacement. This is in part due to the fact that at $z=0$ the $P_{\rm GRH}/P_{\rm nlDGP} $ ratio is larger than at $z=1$ at large scales, so the Vainshtein effect has to overcome a larger difference at $z=0$. A Vainshtein scale (analogous to $r_*$ in the Schwarzshild case)  may be defined by the scale at which $P_{\rm GRH}/P_{\rm nlDGP} $  starts to decrease, this is about $k_*\simeq 2 \kvecMpc$ at $z=1$ and $k_*\simeq 1 \kvecMpc$ at $z=0$.  Note that at intermediate scales the $P_{\rm GRH}/P_{\rm nlDGP} $ goes up slightly, this is expected as the $P_{\rm GRH}$ is enhanced by nonlinearities more than $P_{\rm nlDGP} $  due to its larger growth factor or $\sigma_8$.

The solid lines in Figs.~\ref{fig:pk_z1} and~\ref{fig:pk_z0} denote the predictions from paper I for $P_{\rm nlDGP}$ (left panel) and $P_{\rm GRH}/P_{\rm nlDGP}$ (right panel). We see very good agreement, at the few percent level for the ratios, and slightly worse for the absolute power (about $5,3\%$ for $z=0,1$, respectively, at quasilinear scales), until smoothing due to the CIC kernel substantially cuts the power at high-$k$. The prediction tends to underestimate the N-body simulations, and this is expected to some extent from the details described in paper~I. First, we neglected the extra mode-coupling induced in modified gravity, which was estimated in paper~I to be about 1-2\% from the bispectrum results. Second, the predictions use the GR HaloFit fitting formula~\cite{HF}, which is known to underestimate the nonlinear power on quasilinear scales by up to 4,2\% at $z=0,1$~\cite{RPTbao}. Therefore, we consider these results to be more than reasonable agreement, especially since there is no free parameter being fit in making these predictions (see paper~I).

\begin{figure}[!t]
\centering
\includegraphics[angle=0, width=\linewidth]{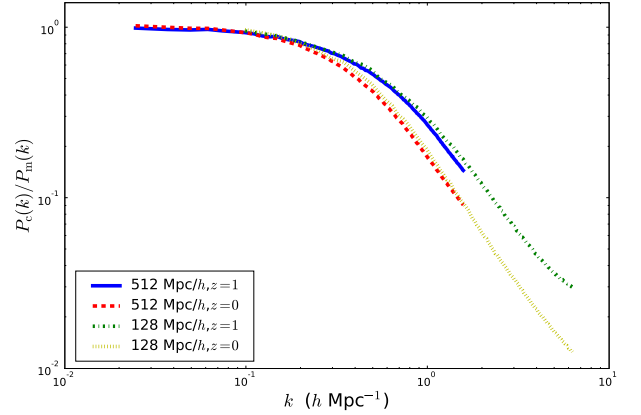}
\caption{ The ratio of power spectrum of the normalized brane-bending mode $c$ (see Eq.~\protect\ref{cdef}), $P_c$, to the nonlinear
density  power spectrum $P_{\rm m}$. We show $P_{\rm c}$ from the 512 and 128
 Mpc/$h$ simulations at $z=0$ and 1.  At small $k$ the ratio is close to 1, but $c$ is increasingly suppressed at high $k$. The suppression is also more pronounced for lower redshift, as expected. }
\label{fig:cpk_ratio}
\end{figure}

To see the Vainshtein effect in more detail it is useful to check the suppression of the brane extrinsic curvature represented by $\nabla^2 C$. Note that if in Eq.~(\ref{eq:FullDGP2}) the nonlinear terms and non-local terms are neglected, we have the linearized relation
\begin{equation}
\delta = \frac{\alpha  \eta }{3 ( \eta - 1) } \bar{\nabla}^2 C.
\label{deltaClin}
\end{equation}
Thus  it is convenient to define
\begin{equation}
c \equiv \frac{\alpha  \eta }{3 ( \eta - 1) } \bar{\nabla}^2 C,
\label{cdef}
\end{equation}
and study the suppression of $c$ with respect to $\delta$ expected from the Vainshtein mechanism at small scales. In Figure~\ref{fig:cpk_ratio}, we plot the ratio of the power spectrum of $c$, $P_{\rm c}$, and the nonlinear density power spectrum $P_{\rm m}$. Again
we see that the ratio is 1 in small $k$ limit, whereas it approaches zero  as $k$ increases. The suppression is more appreciable at $z$=0 than $z$=1, as expected.

\begin{figure}[!ht]
\centering
\includegraphics[angle=0, width=\linewidth]{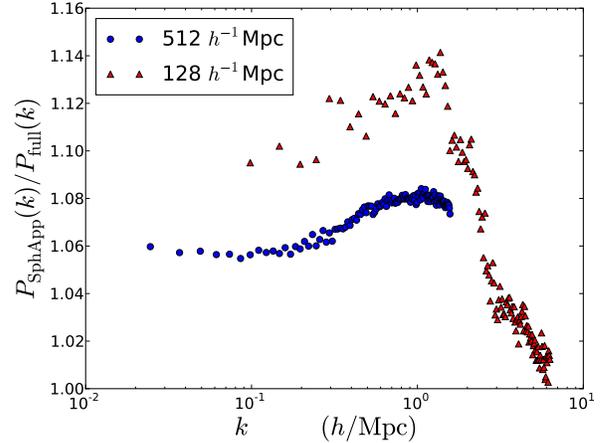}
\caption{Ratio of $P_{\rm nlDGP}$ in the spherical approximation to the full solution as a function of scale for a simulation in a box of  size $512 \Mpc$ (blue circles) and $128 \Mpc$ (red triangles) at $z=0$.}
\label{fig:spherical}
\end{figure}

Given the complexities of simulating the nonlinear PDE for $C$, Eq.~(\ref{eq:FullDGP2}), one may want to resort to approximations, and the spherical approximation immediately comes to mind. In this case one can convert Eq.~(\ref{eq:FullDGP2}) into a local equation in terms of $\nabla^2C$ at small scales. Indeed, spherical symmetry means that $(\nabla_{ij} C)^2 = (1/3) (\nabla^2C)^2$, i.e. this corresponds to spherically average the Fourier space kernel in Eq.~(\ref{eq:NLFourier}) over $\mu$, giving 

\begin{eqnarray}
\label{eq:FullDGP2sph}
{2\over 3} (\bar{\nabla}^2 C)^2   &  + &   \alpha  \bar{\nabla}^2 C  +    \frac{ 3 \beta (\eta -1) }{2 \eta
-1 }    \sqrt{ - \bar{\nabla}^2  } C              \nonumber  \\ 
     & = & \frac{ 3( \eta -1 ) } {\eta } ( 1
- \beta \bar{\nabla}^{-1} ) \delta,  
\end{eqnarray}
which can be solved very easily at small scales (where nonlocal terms can be neglected), leading to the following modified Poisson equation~\cite{LueSS,paper1}
\beq
\bar{\nabla}^2 \phi = {3\over 2}\frac{\eta-1}{\eta} \delta - {1\over 2\eta}\Big( {3\alpha\over 4}\Big) 
\Big[ \sqrt{1+{8\over \alpha^2}\frac{\eta-1}{\eta} \delta}-1\Big].
\label{SphPoi}
\eeq
In practice we do include the nonlocal linear terms (although they are negligible for the box size we will show results for); then solving Eq.~(\ref{SphPoi}) is equivalent to simply using the first iteration of our FFT-relaxation algorithm. Indeed, for our choice of $w=1/3$, the first step corresponds to neglecting the quadrupole in the nonlinear nonlocal terms and solve for the monopole. This is similar to the method used in~\cite{Khoury} to approximate the nonlinear Poisson equation for the DGP model and more generic degravitation models.

Figure~\ref{fig:spherical} shows the results of running our code in spherical mode compared to the full solution. Overall we see good agreement, at approximately the 10\% level. However, there are important differences to note. First, there is disagreement between the two box sizes used ($128\Mpc$ and $512\Mpc$); second, the large-scale power of the two simulations differ among themselves and from the full solution. The reason for these results  can be understood as follows. Making the spherical approximation corresponds to replacing $(1-\mu^2) \to 2/3$ in the Fourier space kernel in Eq.~(\ref{eq:NLFourier}), and that means there is considerable back reaction on linear modes from small scales. Indeed, defining
\beq
\Phi(\k) \equiv \int  d^3k_1d^3k_2   
\, [ \delta_{\rm D}]\,  [1- \mu^2]\, k_1^2C(\bold{k}_1)\, k_2^2C(\bold{k}_2),
\label{Phi}
\eeq
where $[\delta_{\rm D}]\equiv  \delta_{\rm D}(\bold{k}-\bold{k}_1-\bold{k}_2)$, we are interested in estimating how much such a term would contribute to the growth factor at large scales. To do so, it is convenient to cross correlate $\Phi$ with the large scale density perturbation,
\beqa
\lexp \Phi(\k) \delta(\k') \rexp & =&\delta_{\rm D}(\k+\k') \  \Big[  \frac{3(\eta-1)}{\alpha\eta}\Big]^2 
 \nonumber \\
&  \times & \int  d^3q\, [1- \mu^2]\, B(\k-\q,\q)
\label{PhiCross}
\eeqa
where we used Eq.~(\ref{deltaClin}) to relate $\nabla^2C$ to $\delta$ at large scales, $\mu$ is now the cosine of the angle between $\k-\q$ and $\q$, and $B$ denotes the bispectrum of density perturbations. Now as $k\to 0$, we have
\beq
(1-\mu^2)\to {k^2\over q^2} (1-\mu_{kq}^2)
\label{supp}
\eeq
where $\mu_{kq}$ is the cosine of the angle between $\k$ and $\q$. Note that this geometric factor suppresses the contribution of the nonlinear term to large scale modes. We are interested in the terms proportional to the large scale power spectrum, as we are looking effectively for the contribution in $\Phi(\k)$ proportional to $\delta(\k)$. Using the second-order perturbative expression for the bispectrum we have that as $\k\to 0$ such contribution reads
\beq
\lexp \Phi\, \delta \rexp \propto k^2 \sigma_v^2\, P(k),
\label{LowK}
\eeq
where $\sigma_v^2 = \int d^3 q P(q)/q^2$. We thus see that in the full theory the contribution of these terms to large scales is suppressed by the factor $k^2 \sigma_v^2$ at large scales, ensuring the validity of linear perturbation theory at sufficiently large scales.

However, things change drastically if we take the spherical approximation $(1-\mu^2) \to 2/3$, because the suppression factor from Eq.~(\ref{supp}) is no longer present, leading to instead
\beq
\lexp \Phi_{\rm spherical}\, \delta \rexp \propto  \sigma^2\, P(k),
\label{LowK2}
\eeq
where $\sigma^2 = \int d^3 q P(q)$. We thus have a contribution to large scale modes which is formally divergent, proportional to the variance of the density field and not suppressed by $k$. That's why the growth factor at large scales is more affected for the smaller simulation box ($128\Mpc$) where there are smaller scale modes contributing to a larger $\sigma^2$ than for the $512\Mpc$ box. We also checked that the correction to the linear growth factor is smaller at higher redshift where $\sigma^2$ is smaller than at $z=0$. 

We conclude that the spherical approximation is reasonable at the 10\% level but one should be careful that the solution is affected at all scales (including linear). From Fig.~\ref{fig:spherical} we see that the Vainshtein effect is a bit stronger in the spherical approximation; this can be understood from a calculation of the parameter $Q_{\rm eff}$ in paper~I, replacing the kernel $(1-\mu^2)$ by a constant increases $Q_{\rm eff}$. That's why despite having larger power than linear the spherical model approximates the full solution at the smallest scales (see Fig.~\ref{fig:spherical}).

\begin{figure}[!t]
\centering
\includegraphics[angle=0, width=\linewidth]{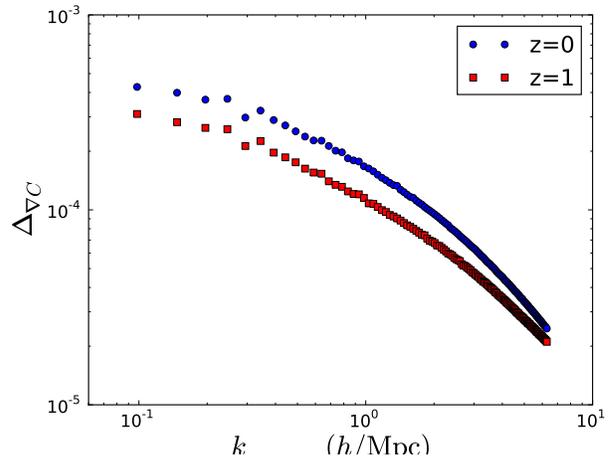}
\caption{Dimensionless power spectrum of brane-bending metric perturbations $\Delta_{\nabla C}$ (see Eq.~\protect\ref{DGradC}) as a function of scale at $z=0,1$. Despite the fact that the dynamics of $C$ is nonlinear, metric perturbations remain in the weak field limit.}
\label{fig:gi4}
\end{figure}

As the onset of the Vainshtein mechanism is related to nonlinearities in the dynamics of $C$ it is worth checking to what extent the metric perturbations related to $C$ are in the weak field limit or not. As discussed in paper~I, while nonlinearities in the dynamics of $C$ are strong (meaning the extrinsic curvature fluctuations $\delta K/K$ become larger than unity), the metric perturbations should remain small throughout evolution. To check this, we construct the dimensionless power spectrum of the metric coefficient (brane-bending mode) related to $C$, $g_{4i}=\nabla_i C$, i.e.
\beq
\Delta_{\nabla C}(k) \equiv 4\pi k^3 P_{\nabla C}(k)= 4\pi k^5 P_C(k).
\label{DGradC}
\eeq
Figure~\ref{fig:gi4} shows $\Delta_{\nabla C}$ as a function of $k$ for $z=0,1$, confirming that in fact $\Delta_{\nabla C}$ {\em decreases} as $k$ increases. There are two reasons for this: first, the Vainshtein mechanism suppresses $C$ in the nonlinear regime; second,  extrinsic curvature fluctuations $\delta K/K\sim (k/aH) (k/a)C$ go nonlinear at scales where $(k/aH) \gg 1$. From these numerical results we conclude that the Vainshtein mechanism works as expected, and a cosmological background of perturbations is enough to suppress brane-bending mode fluctuations significantly at  Mpc scales.

\begin{figure}[!t]
\centering
\includegraphics[angle=0, width=\linewidth]{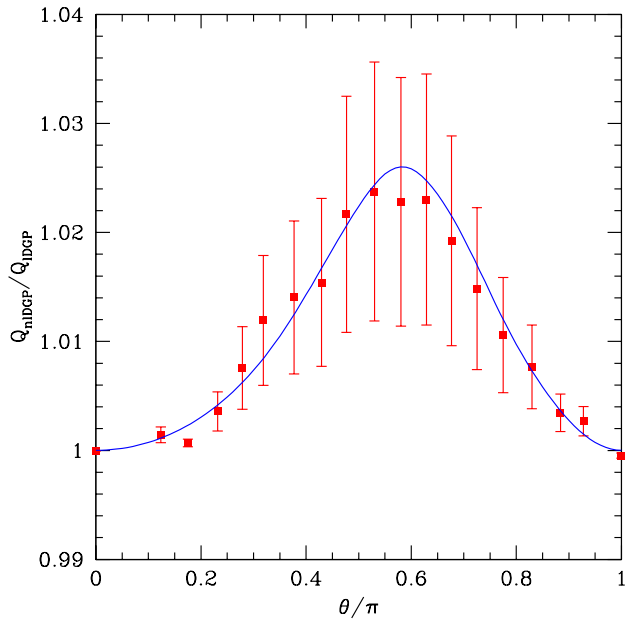}
\caption{Ratio of bispectrum (nonlinear divided by linear DGP) for triangles with $k_1=0.1\kvecMpc$ and $k_2=0.2\kvecMpc$ as a function of angle $\theta$ between $\k_1$ and $\k_2$. Solid lines denote the perturbative prediction from paper~I, symbols with error bars denote measurements in N-Body simulations.}
\label{fig:Bisp}
\end{figure}

\section{The Bispectrum}
\label{sec:Bispectrum}

The nonlinearities in the modified gravity sector (dynamics of $C$) leave signatures in the observed non-Gaussian properties of density perturbations, as discussed in detail in paper~I. The prediction~\cite{paper1} is that the density field bispectrum should show an enhancement for isosceles triangles and no difference for squeezed triangles (where the nonlinear kernel in the modified gravity sector, Eq.~(\ref{eq:NLFourier}), vanishes since $\mu =\pm 1$). 

Comparing the bispectrum in nonlinear DGP simulations versus standard gravity simulations is difficult because they have different power spectrum amplitudes at low redshift, due to the difference in growth factors. As a result, the standard (i.e. present in GR) nonlinear effects will be slightly different and can mask the nonlinear effects we are interested in here if those differences are not taken into account accurately enough. The best way to see the bispectrum enhancement is to compare nonlinear and linearized DGP simulations, which only differ in the presence of the nonlinearities in the modified gravity sector.

Figure~\ref{fig:Bisp} shows such a comparison for triangles with $k_1=0.1\kvecMpc$ and $k_2=0.2\kvecMpc$ as a function of angle $\theta$ between $\k_1$ and $\k_2$. What we compare is the ratio of the reduced bispectra, defined as
\beq
Q_{123} \equiv {B_{123} \over (P_1P_2+P_2P_3+P_3P_1)},
\label{Qdef}
\eeq
where $B$ denotes the bispectrum and subindices label wavectors, e.g. $P_i=P(k_i)$. 
The solid lines in Fig.~\ref{fig:Bisp} show the expected ratio from perturbation theory as calculated in paper~I (note however that $\Omega_m^{0}=0.27$ here instead of 0.2), while symbols with error bars denote measurements in our $N$-body simulations (4 realizations with box size $1280 \Mpc$) for linear and nonlinear DGP models. The very good agreement is another (nontrivial) check of our numerical code.

\begin{figure*}[t!]
\begin{center}
\includegraphics[width=0.49\textwidth]{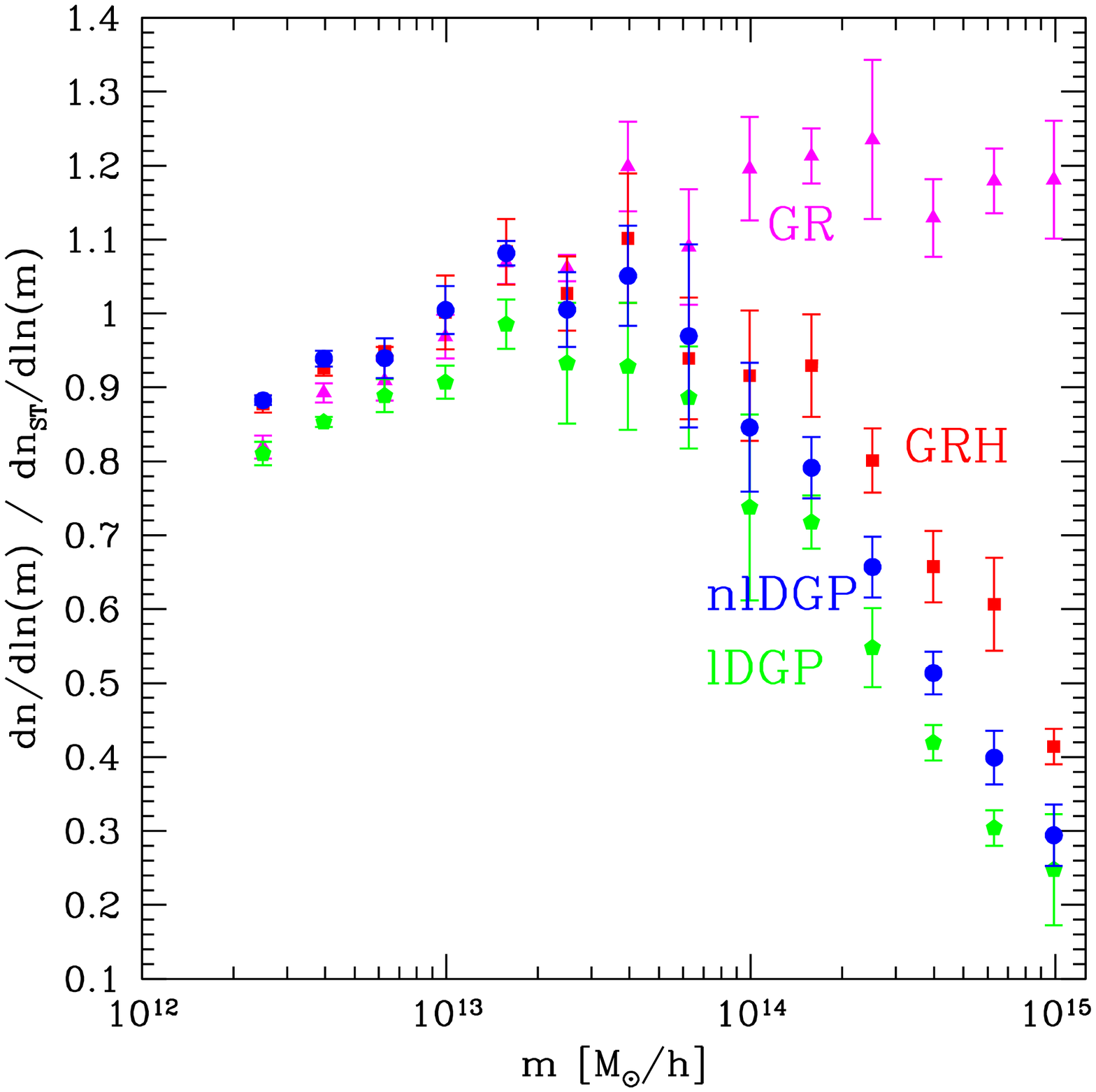}
\includegraphics[width=0.49\textwidth]{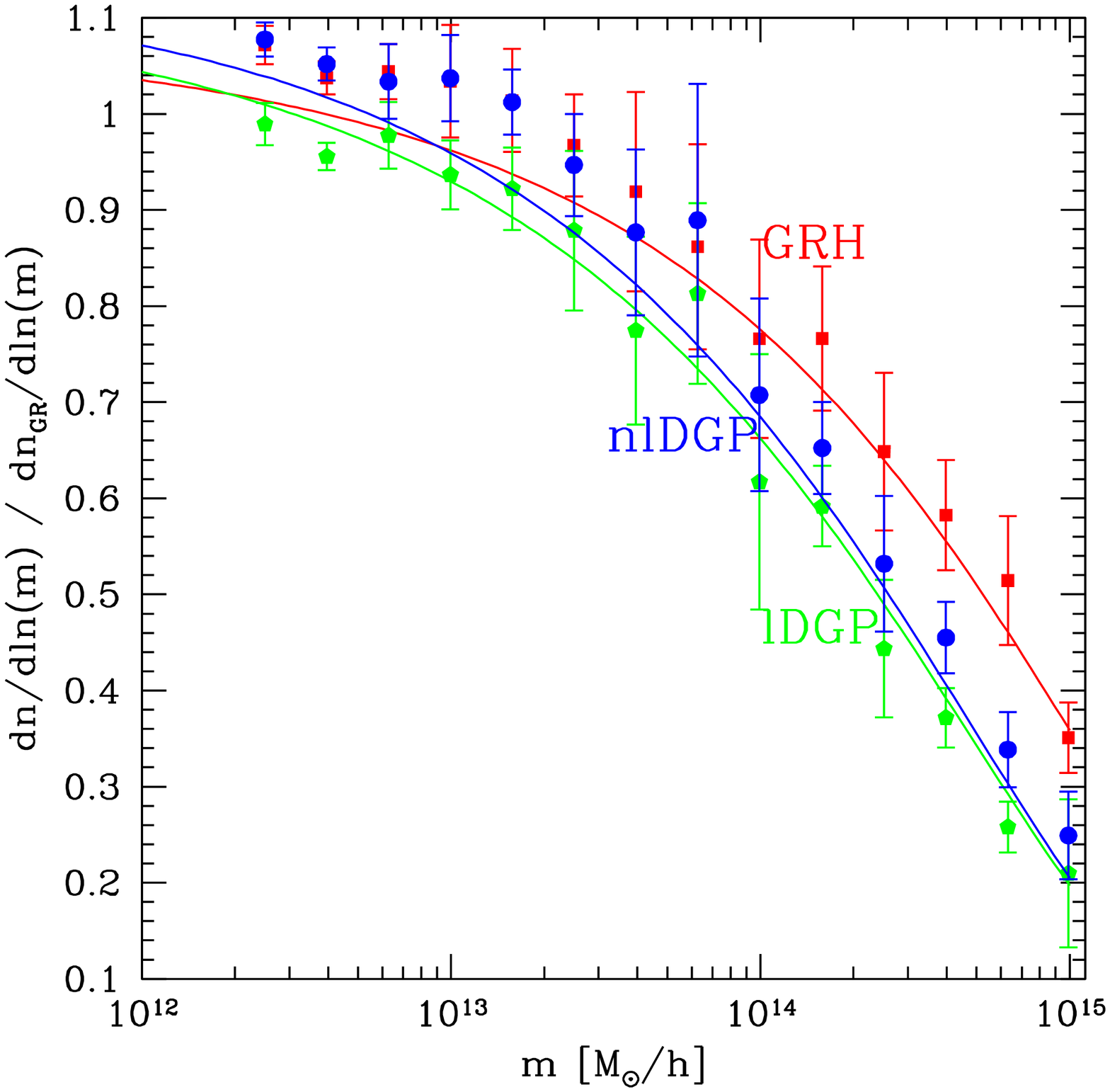} 
\caption{ {\em Left panel:} mass functions measured in numerical simulations for the different models, as labeled, divided by a reference ST mass function. {\em Right panel:} the measured mass function for the three non standard models divided by the GR mass function from the simulations. The solid lines denote the predictions for the ST mass function ratios based on the renormalized linear spectra calculated in Paper I.  }
\label{fig:massfn}
\end{center}
\end{figure*}
\section{Dark Matter Halos}
\label{sec:Halos}

\subsection{Mass Function}
\label{sec:MassFunction}

We identify dark matter halos from the simulations using the Friend-Of-Friend algorithm \cite{Davis} with linking length equal to 0.2 times the interparticle separation. We construct the mass function $(d n /d \ln M)$, the number density of halos per unit comoving volume per unit interval of $\ln M$, for halos that have more than 20 particles. We use box sizes of $128 \Mpc$, $256 \Mpc$ and $512 \Mpc$. We run 4 simulations for each box size and model. 

To better see the differences among the models we run, we divide our simulation results by the Sheth-Tormen (hereafter ST,~\cite{ShethTormen}) mass function for the GR case
\begin{equation}
\label{eq:STfun}
{dn_{\rm ST} \over d \ln M} = \frac{\bar{\rho}_{\rm m} }{M} \frac{d \ln \nu } {d \ln M} \nu f(\nu)
\end{equation}
where $\bar{\rho}_{\rm m}$ is the background density of matter, and $\nu=(\delta_{\rm c} / \sigma)^2$, with $\delta_{\rm c}\approx 1.68$ the linear overdensity above which a region will collapse according to the spherical collapse model, and $\sigma^2$ is the variance of density fluctuations smoothed with a top-hat filter (with radius related to halo mass by $M = (4 \pi/3) \bar{\rho}_{\rm m } R^3 $). The function $\nu f(\nu)$ is a generalization of the standard Press-Schechter function \cite{PressSchechter} motivated by the ellipsoidal collapse model and free parameters chosen to fit the numerical simulations, and is given by 
\begin{equation}
\nu f(\nu) = \frac{A}{\sqrt{\pi} } ( 1 + \frac{1 }{\nu'^p }) \left(
 \frac{\nu'}{2 }   \right)^{  \frac{1}{2}  }    e^{-\nu' / 2 },    
\end{equation}
where $\nu' = a \nu $, with $a =0.707$. The numerical values of $A$ and $p$ are 0.322 and 0.3 respectively.

The left panel of Fig.~\ref{fig:massfn} shows the measured mass functions from the numerical simulations for the different models normalized as discussed above. To minimize the effects of poor mass resolution we use results from each box size for halo that contain more than about 300 particles; when more than one box contributes to a given bin in mass, we average results from different box sizes weighted by the volume of the corresponding boxes. Still, in the low mass regime ($M\la10^{12.5}h^{-1}M_\odot$), the differences for the GR case from the ST mass function are most likely due to poor mass resolution, but again we expect the relative differences among different models (which are very small) to be an accurate representation of what would happen in higher resolution runs. At high mass ($M\ga 10^{13.5}h^{-1}M_\odot$), the enhancement for the GR case over the ST mass function is consistent with what is seen using Gadget (see  e.g.~\cite{2lpt}), and it is in this regime where the deviations among the different models is largest.

The behavior seen at the high mass end is as expected, since the GR model has the largest $\sigma_8$, followed by the GRH model, and then lDGP and nlDGP models that share the same linear normalization. The difference between lDGP and nlDGP models is reasonable given the Vainshtein effect, which makes the power spectrum of the nlDGP case larger than lDGP. At low mass, the differences are much smaller, similar to what happens with GR models of different cosmological parameters where the mass function does not evolve much for $M\ll M_*$. In this regime the nlDGP model has a larger mass function than the lDGP model, again this is expected from the Vainshtein mechanism, and indeed the nlDGP model agrees with the GRH model at small mass, as expected from the matching of their power spectra at small scales.

Motivated by these results, the right panel of Fig.~\ref{fig:massfn} shows the mass function calculated from Eq.~(\ref{eq:STfun}) using the renormalized linear power spectra as calculated in paper~I. Since $\delta_{\rm c } =1.68$ works very well for different cosmologies in the case of GR (e.g.~\cite{Jenkins}), we take this value for all models we consider regardless whether gravity is modified or not. Corrections to this in the DGP model are expected to be small~\cite{LueSS}. While we do not expect this mapping to work in too much detail, the idea is that we are including the Vainshtein mechanism in the renormalized linear spectra for nlDGP derived in paper~I and using the ellipsoidal collapse model assumed in the ST mass function to take into account the standard gravity nonlinearities. For work on spherical collapse model implications for the mass function in modified gravity models see~\cite{MSS08}. The predictions of paper~I do reflect roughly the features seen in the measurements. Indeed, the relative ordering of the mass function measurements agree with expectations based on the predictions.

\begin{figure}[!t]
\centering
\includegraphics[angle=0, width=\linewidth]{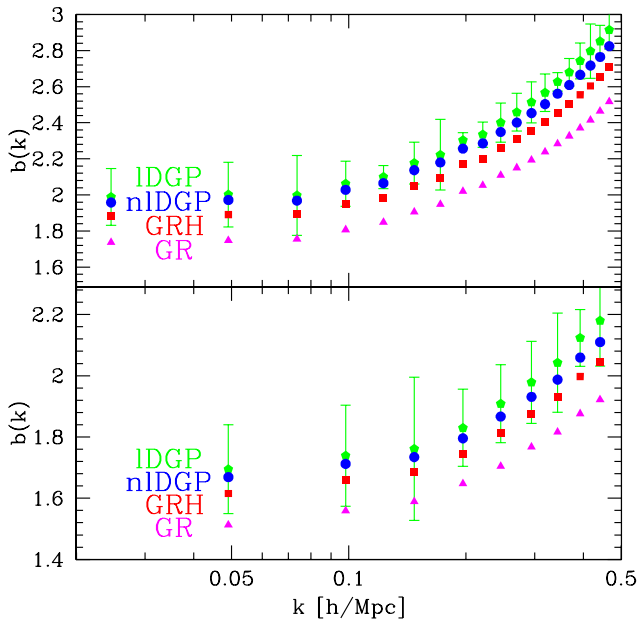}
\caption{The halo bias for halos with mass $M>1.5 \times 10^{12}M_\odot/h$ (top four curves) and $M>1.8 \times 10^{11}M_\odot/h$ (bottom four curves) for different models.  }
\label{fig:halobias}
\end{figure}

\subsection{ Halo Bias}
\label{sec:Bias}

We now discuss the difference in the large-scale bias of dark matter halos for the different models discussed above. Due to our relatively low particle number our mass resolution is limited, therefore we simply concentrate on all halos with more than 20 particles to compute the bias.  The halo bias is obtained from the ratio of the cross halo-matter power spectrum and the dark matter power spectrum, to avoid accounting from possibly nontrivial shot noise in the halo-halo power spectrum. Figure~\ref{fig:halobias} shows the halo bias from simulations of
box size $256 \Mpc$ and $128 \Mpc$, with the same threshold of 20 particles per halo this corresponds to $b(M>1.5 \times 10^{12}M_\odot/h)$ and  $b(M>1.8\times 10^{11}M_\odot/h)$, respectively.  The error bars correspond to the mean over four realizations.

At fixed mass threshold, the relative magnitude of the bias agrees with expectations based on the peak-background split~\cite{BBKS,CK}, which says that the bias is proportional to the slope of the mass function. One can check that the ordering of the bias factors in Fig.~\ref{fig:halobias} at fixed threshold mass qualitatively agrees with this simple prescription by looking at the slopes of the mass functions in the left and right panels of Fig.~\ref{fig:massfn}.

In \cite{Hui} it is pointed out that a scale dependence of the growth factor (as it happens in DGP models at large scales due to nonlocal terms) can give rise to a scale dependence in the halo (or galaxy) bias. Assuming no merging and that halos move with unbiased velocities, the  halo bias $b(z,\bold{k})$ in Fourier space evolves as~\cite{Hui} 
\begin{equation}
b(z,\bold{k}) = 1 + (b_{0} - 1 ) \frac{ \delta_{\rm m }(z_0, \bold{k} ) }{ \delta_{\rm m }(z,\bold{k} ) },  
\label{HBSD} 
\end{equation}
where  $b_0$ denotes the halo bias at formation time, and $\delta_{\rm m }$ is the density contrast of the matter. Even if $b_0$ is independent of scale, the ratio of matter density fluctuations at different times is expected to show some scale dependence due to nonlocal terms in the modified Poisson equation (see paper~I). However, we were unable to test this prediction because 1) the scale dependence in the growth factor is small and kicks in at large scales, 2) to probe large scales, large boxes are required, and our limits on computational resources required means we cannot use large number of particles, which leads to poor mass resolution, 3) for large halo masses (poor mass resolution), halos form late and there is little time between formation and $z=0$ to see a relative scale dependence induced by the ratio of matter fluctuations in Eq.~(\ref{HBSD}). As a result of these factors, the predicted scale dependence is too small and consistent with the uncertainties from our simulations.

\section{Conclusions}
\label{sec:Concl}

We developed a numerical algorithm, which we call FFT-relaxation, to solve the nonlinear field equations (Eqs.~\ref{eq:FullDGP1}-\ref{eq:FullDGP2}) in the DGP model. This enabled us to run fully nonlinear $N$-body simulations of this model and compare to simulations where the field equations are linearized, where standard gravity is used only for perturbations (but not for the background), and to fully standard gravity simulations. Our numerical algorithm includes not only the new nonlinear couplings in the modified gravity sector but also linear but nonlocal terms that lead to 5D behavior at large scales in the quasistatic approximation. We expect that our implementation may be  useful in other scenarios of brane-induced gravity or massive-gravity related models, e.g. \cite{deRham1,deRham2,Nicolis}, which have both of these features.  We extracted the power spectrum, bispectrum, mass function and halo bias from the $N$-body simulations.

From the simulation results, we confirm the expectations based on analytic arguments that the Vainshtein mechanism does operate as anticipated, with the density power spectrum approaching that of standard gravity within a modified background evolution in the nonlinear regime. The transition is very broad and there is no well defined Vainshtein scale, but roughly this corresponds to wavenumbers $k_*\simeq 2 \kvecMpc$ at redshift $z=1$ and $k_*\simeq 1 \kvecMpc$ at $z=0$. We checked that while extrinsic curvature fluctuations go nonlinear, and the dynamics of the brane-bending mode $C$ receives important nonlinear corrections, this extra scalar mode does get suppressed compared to density perturbations, effectively decoupling from the standard gravity sector. At the same time, there is no violation of the weak field limit for metric perturbations associated with $C$. We also quantified the use of the spherical approximation to the nonlinear dynamics of $C$ (see e.g.~\cite{Khoury} for application of this to more generic models), and showed that the results on the density power spectrum show differences of order 10\% at all scales, depending on box size and resolution. The differences are not restricted to small scales as the approximation introduces a spurious enhanced coupling of large-scale modes to small scale modes.

In the nonlinear regime, modified gravity models become difficult to distinguish from standard gravity for a fixed expansion history, and therefore this is not the best place to look for modifications of gravity in observations.  The best way to test these models is on weakly nonlinear scales, we showed that the bispectrum in our simulation presents a small signature of modified gravity due to the additional nonlinear couplings, as predicted by perturbation theory in paper~I; however, this is too small to be tested in present observations. We also studied the abundance and clustering of dark matter halos and found that the largest differences in the mass function are seen at large masses, as anticipated from the results in the power spectrum. The differences seen in the halo bias are consistent with expectations based on the peak-background split.

We tested the calculations presented in paper~I that rely on a renormalization of the linear spectrum due to nonlinear effects in the modified gravity sector, which are induced by the Vainshtein mechanism and modify the effective gravitational constant from linear theory. The predictions for the nonlinear spectrum are in very reasonable agreement with our simulations. A similar prediction for the mass function shows the right trends although our tests are modest due to limited simulation volume and mass resolution. We hope to improve on these tests to sharpen theoretical predictions of modified gravity in the near future. 

Our results suggest that the framework laid down in paper~I, where the nonlinear effects of gravity beyond general relativity are included in the running of the gravitational constant, is quite accurate. As discussed in paper~I, this is in fact a result of the fact that modifications beyond such running (such as extra contributions to non-Gaussianity) are higher-order in $\delta G/G$ and are highly suppressed at small scales. This general argument should be applicable to other theories of gravity, leading to a powerful way to test gravity theories against observations.

\vskip 2pc

While this paper and its companion paper~I~\cite{paper1} were slowly written down, preprint~\cite{SchmidtNB} appeared on arXiv, where N-body simulations of the  DGP model are also presented. His results on the power spectrum and halo mass function are broadly consistent with ours.

\acknowledgements 

We thank C.~Deffayet, G.~Gabadadze, A.~MacFadyen and F.~Schmidt for discussions and X. Feng for useful comments on the Monge-Ampere equation. We also thank the participants of the workshop on Consistent Infrared Modification of Gravity (Paris, November 2008), where this work was first presented, for feedback.  Part of the computational resources for this work were done on NYU computing clusters, we thank Joseph Hargitai and NYU ITS for support. This work was partially supported by grants NSF AST-0607747 and NASA NNG06GH21G. R.~S. thanks the Aspen Center for Physics where this work was completed.

\appendix

\section{Other Aspects of the field equations}
\label{sec:GS-relax}
In this appendix, we would like to compare the FFT-relaxation method with more common relaxation methods. Since the difficulty in solving the Eq.~(\ref{eq:FullDGP2}) stems from the nonlinear terms and it is difficult to implement the nonlocal terms in the usual finite differencing schemes, in this Appendix we discard the nonlocal terms for simplicity. Thus here we are interested in solving the simplified equation
\begin{equation}
\label{eq:FullDGP2simp}
( \bar{\nabla}^2 C )^2 + \alpha  \bar{\nabla}^2 C   - (\bar{
 \nabla}_{ij} C)^2   =  \frac{ 3( \eta -1 ) } {\eta }  \delta.   
\end{equation}

\subsection{Ellipticity of the Equation}  

Before attempting to solve Eq.~(\ref{eq:FullDGP2simp}), some comments are in order. First, Eq.~(\ref{eq:FullDGP2simp}) belongs to the class of {\em fully nonlinear partial differential equations} (PDE), since it is nonlinear in the highest derivative of the unknown. Fully nonlinear PDEs are usually more difficult to deal with than those nonlinear in the field, such as the field equation in $f(R)$ models \cite{Oyaizu1} or nonlinear in lower derivatives of the field. So far there is a general solution theory for this kind of fully nonlinear PDEs only if Eq.~(\ref{eq:FullDGP2simp}) is elliptic in the whole domain \cite{Feng}; however, to determine whether Eq.~(\ref{eq:FullDGP2simp}) is elliptic or not is nontrivial. 

In general for a PDE $F(C)=0$, in our case
\begin{equation}
\label{eq:F(C)}
F(C)= ( \bar{\nabla}^2 C )^2 + \alpha  \bar{\nabla}^2 C   - (\bar{
 \nabla}_{ij} C)^2   -  \frac{ 3( \eta -1 ) } {\eta }  \delta,
\end{equation}
one needs to solve for the eigenvalues of the matrix
\begin{eqnarray}
\begin{pmatrix}
 F_{\bar{\nabla}_{xx}C } &  F_{\bar{\nabla}_{xy}C }  &  F_{\bar{\nabla}_{xz}C }   \\
  F_{\bar{\nabla}_{yx}C } &  F_{\bar{\nabla}_{yy}C } &  F_{\bar{\nabla}_{yz}C }  \\
  F_{\bar{\nabla}_{zx}C } &  F_{\bar{\nabla}_{zy}C } &  F_{\bar{\nabla}_{zz}C }  \\
\end{pmatrix}     ,   
\label{eq:EllipticMatrix}
\end{eqnarray}
where $ F_{\bar{\nabla}_{xx}C }$ denotes partial differentiation of $F$ with respect to $\bar{\nabla}_{xx}C  $, etc. The equation $F(C)=0$ is elliptic if all the eigenvalues are of the same sign, hyperbolic if the sign of one of the eigenvalue differs from the rest, and parabolic if one of the eigenvalues is zero.

However, in 3D, the resultant equation depends on the field $C$, and we cannot determine its nature of the equation without knowing $C$! Nonetheless, we can check the ellipticity of the equation using the solution we obtained from the FFT-relaxation method. We do this test in a $64 \Mpc$ simulation box using $256^3$ particles. We input the convergent solution $C(\bold{x})$ into the matrix Eq.~(\ref{eq:EllipticMatrix}) and solve for its eigenvalues at every grid point. We then check if the signs of the eigenvalues are the same or not. We find that the equation is elliptic at every point in the simulation box.  

It is important to note that violation of ellipticity happens when the bad fraction is non-zero. In our test we used $w=1/3 $ and therefore the bad faction is zero all the time. If $w=0$ is used instead, the bad fraction will be non-zero as shown in the upper panel of Fig.~\ref{fig:res_badfrac_1_0_1o3}, and violation of ellipticity (by substantial fraction of points, of order 10\%) will occur in the same time  interval during which the bad fraction is non-zero. Thus satisfying the ellipticity condition is another self-consistent check of the FFT-relaxation method.  

If Eq.~(\ref{eq:FullDGP2simp}) were in 2D rather than 3D, its nonlinear structure would be the same as the well-known Monge-Ampere
equation \cite{Polyanin}. This simplification is possible if we artificially  assume that $\delta$ is constant along the $z$ direction, with $C$ a function of $x$ and $y$ alone. In 2D, Eq.~(\ref{eq:FullDGP2simp}) reduces to 
\begin{equation}
\label{eq:MA}
2 ( \bar{\nabla}_{xx}C  \bar{\nabla}_{yy} C  - ( \bar{\nabla}_{xy}C)^2  ) +
\alpha ( \bar{\nabla}_{xx}C + \bar{\nabla}_{yy} C ) = \frac{3(\eta -1)}
{\eta} \delta    
\end{equation}
and the determinant of the corresponding matrix Eq.~(\ref{eq:EllipticMatrix}) in 2D, after making use of Eq.~(\ref{eq:MA}), is
\begin{equation}
\label{eq:Det2D}
\alpha^2 + \frac{6( \eta -1)}{\eta} \delta. 
\end{equation}
One can check that Eq.~(\ref{eq:Det2D}) is always positive even for $\delta=-1$. This shows that the eigenvalues are of the same sign, and thus Eq.~(\ref{eq:MA}) is elliptic.

\subsection{On Gauss-Seidel Relaxation}  

Now we turn to solving Eq.~(\ref{eq:FullDGP2simp}) using a finite difference method. One of the potential methods for solving nonlinear PDEs is the Newton-Gauss-Seidel-relaxation method \cite{NR}, which for our case it reads 
\begin{equation}
C^{n+1}_{ijk} = C^n_{ijk} - \frac{F(C^n_{ijk}) }{ F'(C^n_{ijk})},
\end{equation}
where $C^n_{ijk}$ denotes the value of $C$ at the grid point $(i,j,k)$ in the $n^{\rm th}$ iteration and  $F'(C^n_{ijk})$ represents the
derivative of $F$ with respect to $C^n_{ijk}$. We find that this method diverges very quickly. This is probably due to the fact that upon discretization, Eq.~(\ref{eq:FullDGP2simp}) is a quadratic equation in the discrete variable $C_{ijk}$, which has two, one or zero real roots. If
the starting trial solution is not good enough, it is impossible for different points to converge to the same branch of solution.

Recently, algorithms were developed by mathematicians
\cite{MathNumericalMA} to attack the Monge-Ampere equation, but most
of them are difficult to implement, albeit more rigorous than the method we developed in this paper. However, equations similar to Eq.~(\ref{eq:MA}) also arise in other fields such as meteorology, where a 2D equation called the ``balance equation'' \cite{Shuman,Meteorologists} used to model the wind flow is of Monge-Ampere type. The approach followed in those papers is simple enough to implement. They used Gauss-Seidel relaxation together with the quadratic formula instead of Newton's method. This chooses the branch of the solution explicitly and has better control over which branch the solution converges to. In what follows, we will discretize Eq.~(\ref{eq:FullDGP2simp}) similarly to how is done in~\cite{Shuman} and generalize it to 3D.

Le us now discuss the details of the calculations.  We first write,
\begin{widetext}
\begin{eqnarray}
&  &  (\bar{\nabla}^2C)^2 - (\bar{\nabla}_{ij}C)^2   \nonumber   \\ 
&=& 2 [ \bar{\nabla}_{xx}C \bar{\nabla}_{yy}C + \bar{\nabla}_{xx}C\bar{\nabla}_{zz}C  + \bar{\nabla}_{yy}C
   \bar{\nabla}_{zz}C]     -  2 [ (\bar{\nabla}_{xy}C)^2  + (\bar{\nabla}_{xz}C)^2 +
   (\bar{\nabla}_{yz}C)^2 ]  \nonumber       \\ 
&=&   \frac{1}{2}   [ \mu^2 - (\bar{\nabla}_{xx}C -
 \bar{\nabla}_{yy}C)^2 +  \nu^2 -  (\bar{\nabla}_{xx}C -
 \bar{\nabla}_{zz}C)^2   +  \sigma^2 -  (\bar{\nabla}_{yy}C -
 \bar{\nabla}_{zz}C)^2   ]  - 2[  (\bar{\nabla}_{xy}C)^2  +  (\bar{\nabla}_{xz}C)^2 +
 (\bar{\nabla}_{yz}C)^2 ],  \nonumber 
\end{eqnarray}
where in the last line we have denoted $\mu = \bar{\nabla}_{xx}C +
\bar{\nabla}_{yy}C$, $\nu =\bar{\nabla}_{xx}C + \bar{\nabla}_{zz}C$, and $\sigma =
\bar{\nabla}_{yy}C + \bar{\nabla}_{zz}C$ respectively. Now Eq.~(\ref{eq:F(C)}) can be expressed as  
\begin{eqnarray}   
F &=&   \frac{\alpha}{2}( \mu + \nu + \sigma ) + \frac{1}{2}(\mu^2 + \nu^2 +
\sigma^2 ) -  \frac{1}{2} [  (\bar{\nabla}_{xx}C -  \bar{\nabla}_{yy}C)^2 + (\bar{\nabla}_{xx}C -
 \bar{\nabla}_{zz}C)^2   +  (\bar{\nabla}_{yy}C -  \bar{\nabla}_{zz}C)^2      ] \nonumber  \\
& &  -  2 [  (\bar{\nabla}_{xy}C)^2  +  (\bar{\nabla}_{xz}C)^2 +  (\bar{\nabla}_{yz}C)^2 ]
- \frac{3(\eta -1)}{\eta} \delta
\nonumber  \\
&=& \frac{1}{2} \left[ (\mu + \frac{\alpha}{2} )^2  + (\nu +
\frac{\alpha}{2} )^2    +  (\sigma + \frac{\alpha}{2} )^2 - \frac{3 \alpha^2}{4}   \right]   - \frac{1}{2} \left[  (\bar{\nabla}_{xx}C -  \bar{\nabla}_{yy}C)^2 + (\bar{\nabla}_{xx}C -
 \bar{\nabla}_{zz}C)^2   +  (\bar{\nabla}_{yy}C -  \bar{\nabla}_{zz}C)^2       \right] \nonumber  \\
& &  -  2 [  (\bar{\nabla}_{xy}C)^2  +  (\bar{\nabla}_{xz}C)^2 +  (\bar{\nabla}_{yz}C)^2 ]
-  \frac{ 3(\eta -1 ) }{ \eta  }\delta.
\end{eqnarray}
The advantage of writing $F$ in this way is that upon discretization, the central term $C_{ijk}$ only appears in the first square bracket. Suppose we write the difference equation as a quadratic equation in $C^n_{ijk} $, and solve for its root to get the new $C^{n+1}_{ijk}
$ (note that we assume that the iteration is Jacobi-like, \textit{i.e.} we use only the old $C^{n}$ to update the central term $C_{ijk}^{n}$; we will relax this assumption later).  Let's denote the residual before the $n^{\rm th}$ iteration be $F^{n}$, and the residual after the $n^{\rm th}$ iteration be $F^{n+1}$. Explicitly, we have
\begin{eqnarray} 
F^n & = & \frac{1}{2} \left[ (\mu^n + \frac{\alpha}{2} )^2  + (\nu^n +
\frac{\alpha}{2} )^2    +  (\sigma^n + \frac{\alpha}{2} )^2 - \frac{3
 \alpha^2}{4}   \right]  + Z(C^n) -  \frac{ 3(\eta -1 ) }{ \eta
}\delta   \\ 
F^{n+1} & = & \frac{1}{2} \left[ (\mu^{n+1} + \frac{\alpha}{2} )^2  + (\nu^{n+1} +
\frac{\alpha}{2} )^2    +  (\sigma^{n+1} + \frac{\alpha}{2} )^2 - \frac{3
 \alpha^2}{4}   \right]  + Z(C^n) -  \frac{ 3(\eta -1 ) }{ \eta
}\delta ,         
\end{eqnarray}
where we have denoted the non-central terms $Z(C^n)$, which are unaltered in the iteration. Subtracting $F^{n}$ from  $F^{n+1}$ we have 
\begin{eqnarray}
\frac{1}{2} [ (\mu^{n +1} + \frac{\alpha}{2} )^2 +  (\nu^{n +1} +
 \frac{\alpha}{2} )^2  +   (\sigma^{n +1} + \frac{\alpha}{2} )^2    
  -   ( \mu^{n} + \frac{\alpha}{2} )^2 -  ( \nu^{n} +
 \frac{\alpha}{2} )^2   -  ( \sigma^{n } + \frac{\alpha}{2} )^2     ] 
  =  F^{n + 1} - F^{n} . 
\end{eqnarray}

Next, we discretize the variables and the derivative operators, and write $C_{ijk}^{n +1}  $ as  $C_{ijk}^{n} + \xi_{ijk}^n$. Then, for example, we can express
\begin{eqnarray}
\mu^{n+1}_{ijk}  & = & \frac{ C_{i+1,jk}^n + C_{i-1,jk}^n +
 C_{i,j+1,k}^n +C_{i,j-1,k}^n -4 C_{ijk}^{n+1}  }{ \bar{h}^2 }     \\ \nonumber 
&=& \frac{ - 4 } { \bar{h}^2 } \xi_{ijk}^{n} + \mu^n_{ijk}, 
\end{eqnarray}
where $\bar{h}\equiv aHh$ and  $h$ is the size of the grid. The extra factor $aH$ originates from our definition of $\bar{\nabla}$ as $\nabla /(aH)$. Thus we arrive at a quadratic equation in $\xi_{ijk}^n $
\begin{equation}
(\xi_{ijk}^n)^2 - \frac{\bar{h}^2 }{6 } ( \mu_{ijk}^{n} + \nu_{ijk}^{n} +
\sigma_{ijk}^{n} + \frac{3 \alpha }{2}  )  \xi_{ijk}^n - \frac{\bar{h}^4}{24}( F_{ijk}^{n +1} -  F_{ijk}^{n}  )= 0 .
\end{equation}
We cannot solve this equation since $F_{ijk}^{n+1}$ is unknown. The next approximation we make is to set $F_{ijk}^{n+1}$ to zero. This is reasonable as we assume the next iteration will give a smaller residual than the previous one. Thus the final equation we get is 
\begin{equation}
\label{eq:quadratic}
(\xi_{ijk}^n)^2 - \frac{\bar{h}^2 }{6 } ( \mu_{ijk}^{n} + \nu_{ijk}^{n} +
\sigma_{ijk}^{n} + \frac{3 \alpha}{2}  )  \xi_{ijk}^n + \frac{\bar{h}^4}{24}  F_{ijk}^{n} = 0 .
\end{equation}
From Eq.~(\ref{eq:quadratic}), we can solve for $\xi_{ijk}^n $ and then get new $C_{ijk}^{n+1}$. However, the quadratic nature of the equation introduces another complication--- which root to take. Obviously, when we take the limit $F_{ijk}^{n}\to 0$, we should take the root such that $\xi_{ijk}^n\to 0$. Thus we shall take the plus sign of the square root. Again, we may be plagued by bad points since the discriminant may not always be non-negative. Rearranging the terms as in Sec.~\ref{sec:negativedis} may help resolve this issue. We will not, however, pursue further investigation since we will see that this method does not quite converge even when the bad fraction is zero. 

\end{widetext}

So far, we used a Jacobi approach, \textit{i.e.}  in updating $C^{n+1}$ we only use old values of $C^{n}$. Now let's consider a Gauss-Seidel approach, in which we will use the new $C^{n+1}$ whenever they are available during the iteration. We find that this can improve the convergence rate significantly, as it is normally the case. 


\begin{figure}[!t]
\centering
\includegraphics[angle =0, width= \linewidth]{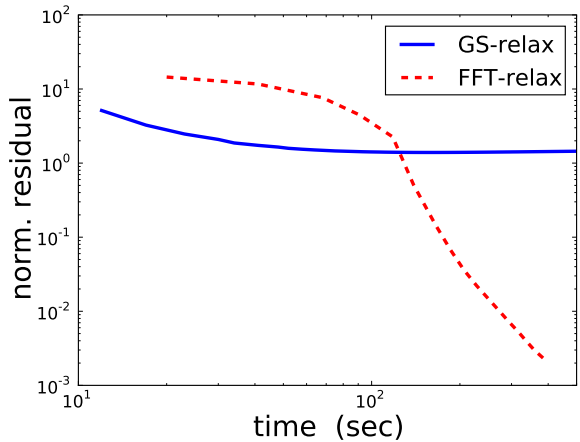}
\caption{Normalized residuals as a function of time for the  FFT-relaxation and GS-relaxation methods. For the GS-relaxation method the normalized residual saturates at some value of order 1, in this case about 1.4, while for the FFT-relaxation method the residual keeps on decreasing.  Note that in practice, we work on the steep part of the FFT-relax curve when we use the previous time step solution to initialize the relaxation method.}
\label{fig:GS_vs_FFT_NormRes}
\end{figure}

We compare the convergence of the GS-relaxation and the FFT-relaxation method in Fig.~\ref{fig:GS_vs_FFT_NormRes}, where we plot the normalized residual versus  time taken for these two methods. We start from the same initial linear solution. We perform this test on a box of $64 \Mpc$ at $z=0$ with $w=1/3$ for FFT-relaxation. In both methods, there are no bad points (in fact for the FFT-relaxation method, a small bad fraction appears in the first few cycles, and then disappears).

We see that although GS-relaxation takes less time to run in each cycle, its normalized residual settles to some value of order 1. In fact the normalized residual rises very slowly as we continue to iterate. On the other hand, for the FFT-relaxation method (although takes longer time for each cycle) the residual keeps on decreasing, especially after the first few cycles. During our simulations, except for the very first time step, we use the previous time step solution as the trial solution for the FFT-relaxation method, which is much better than the linear solution since we then are in the steep-slope regime in Fig.~\ref{fig:GS_vs_FFT_NormRes}. However, it may be useful to combine both methods to achieve greater convergence speed.

It's not clear to us why the GS-relaxation method together with the quadratic formula seems to work in~\cite{Shuman,Meteorologists}, although there are significant differences with our application here. First, the balance equation is 2D rather than 3D. Secondly, the number of grid points considered in~\cite{Shuman,Meteorologists} are only about $10^3$, so the resolution is rather low. Last, and perhaps most importantly, the performance of these nonlinear relaxation methods depend rather sensitively on the values of the parameters such as $\alpha$ and $3(\eta -1)\eta^{-1 } $ and how rough the data $\delta $ is. In cosmology, $\delta$ is not very regular. To summarize, we presented a simple discretization scheme for GS-relaxation method, and it fails to solve Eq.~(\ref{eq:FullDGP2simp}) satisfactorily. However, given the complexity of Eq.~(\ref{eq:FullDGP2simp}), we don't rule out the possibility that a more thoughtful discretization scheme would work. In fact, while this paper was being finished, preprint~\cite{SchmidtNB} appeared on the arXiv showing that a convergent method based on Newton's method can be constructed for these field equations based on multigrid techniques.


\begin{thebibliography}{99}
\small



\bibitem{DGP} G. R. Dvali, G. Gabadadze, and M. Porrati, Phys. Lett. B
 \textbf{485}, 208 (2000).

\bibitem{Deffayet}C. Deffayet, Phys. Lett. B \textbf{502}, 199 (2001). 

\bibitem{Lue}A. Lue, Phys. Rept. \textbf{423}, 1 (2006).
\bibitem{Gabadadze} G. Gabadadze, Nucl. Phys. B (Proc. Suppl.) \textbf{171}, 88 (2007). 

\bibitem{Koyama}K. Koyama, Gen. Relativ. Gravit. \textbf{40}, 421 (2008).


\bibitem{LueSS}A. Lue, R. Scoccimarro, and G. D. Starkman, Phys. Rev. D
 \textbf{69}, 124015 (2004). 

\bibitem{KunzSapone} M. Kunz, and D. Sapone, Phys. Rev. Lett. \textbf{98}, 121301 (2007).

\bibitem{paper1}R. Scoccimarro, submitted  to Phys. Rev. D (2009), arXiv:0906.4545

\bibitem{KTH}K. Koyama, A. Taruya, and T. Hiramatsu, arXiv:0902.0618


\bibitem{Stabenau}H. F. Stabenau, and B. Jain, Phys. Rev. D \textbf{74},
 084007 (2006).

\bibitem{Shirata}A. Shirata, Y. Suto, T. Shiromizu, and N. Yoshida, Phys. Rev. D \textbf{76}, 044026 (2007). 

\bibitem{Laszlo}L. Laszlo, and R. Bean, Phys. Rev. D \textbf{77}, 024048
 (2008).



\bibitem{Oyaizu1} H. Oyaizu, arXiv:0807.2449.
\bibitem{Oyaizu2} H. Oyaizu, M. Lima, and W. Hu, arXiv:0807.2462.
\bibitem{Schmidt} F. Schmidt, M. Lima, H. Oyaizu, and W. Hu, arXiv:
0812.0545.  


\bibitem{KoyamaSilva}K. Koyama, and F. Silva, Phys. Rev. D \textbf{75},
 084040 (2007). 

\bibitem{KoyamaMaartens}K. Koyama, and R. Maartens, JCAP \textbf{0601}, 016
 (2006). 


\bibitem{SSH07} I. Sawicki, Y.-S. Song, and W. Hu, Phys. Rev. D
 \textbf{75}, 064002 (2007).

\bibitem{Cardoso} A. Cardoso, K. Koyama, S. S. Seahra, and
 F. P. Silva, Phys. Rev. D \textbf{77}, 083512 (2008). 




\bibitem{vDVZ}H. van Dam, M. Veltman, Nucl. Phys. B \textbf{22}, 397
 (1970); V. I. Zakharov, JETP Lett. \textbf{12}, 312 (1970). 
\bibitem{Vainshtein}A. I. Vainshtein, Phys. Lett. B \textbf{39}, 393
 (1992). 
\bibitem{DKP}T. Damour, I. I. Kogan, A. Papazoglou, Phys. Rev. D \textbf{67}, 064009 (2003). 

\bibitem{Babichev}E. Babichev, C. Deffayet, R. Ziour, arXiv:0901.0393v2. 
\bibitem{Deffayet02}C. Deffayet, G. Dvali, G. Gabadadze, and
 A. I. Vainshtein, Phys. Rev. D \textbf{65}, 044026 (2002).


\bibitem{Gruzinov}A. Gruzinov, New Astron. \textbf{10}, 311 (2005).
\bibitem{Porrati}M. Porrati, Phys. Lett. B \textbf{534}, 209 (2002). 

\bibitem{Tanaka04} T. Tanaka, Phys. Rev. D \textbf{69}, 024001 (2004). 

\bibitem{GaIg05} G. Gabadadze, A. Iglesias, Phys. Rev. D \textbf{72}, 084024 (2005). 



\bibitem{DGPGhosts} M. A. Lutty, M. Porrati, and R. Rattazzi, JHEP
 \textbf{09}, 029 (2004); A. Nicolis, and R. Rattazzi, JHEP
 \textbf{06}, 059 (2004); C. Deffayet, G. Gabadadze, and A. Iglesias,
 JCAP \textbf{0608}, 012 (2006); G. Dvali, New J. Phys. \textbf{8},
 326 (2006); C. Charmousis, R. Gregory, N. Koloper, and A. Padilla, JHEP \textbf{0610}, 066 (2006). 


\bibitem{Song}Y.-S. Song, I. Sawicki, and W. Hu, Phys. Rev. D
 \textbf{75}, 064003 (2007).
\bibitem{Wang}S. Wang, L. Hui, M. May and Z. Haiman, Phys. Rev. D
 \textbf{76}, 063503 (2007). 

\bibitem{Fang}W. Fang \textit{et al.}, arXiv:0808.2208.

\bibitem{SDSSSNIa} J. Sollerman et al., arXiv:0908.4276; H. Lampeitl et al., arXiv:0910.2193

\bibitem{deRham1}C. de Rham, S. Hofmann, J. Khoury, and
 A. J. Tolley, JCAP \textbf{0802}, 011 (2008).
\bibitem{deRham2}C de Rham \textit{et al.},
Phys. Rev. Lett. \textbf{100}, 251603 (2008). 

\bibitem{Nicolis} A. Nicolis, R. Rattazzi, E. Trincherini, arXiv:0811.2197. 


\bibitem{HockneyEastwood} R. W. Hockney, and J. W. Eastwood, \textit{Computer
 Simulations Using Particles}, (Institute of Physics Publishing, Bristol,
 1988).

\bibitem{Padmanabhan}J. S. Bagla, and T. Padmanabhan, arXiv:astro-ph/0411730.

\bibitem{Kravtsov} A. Kravtsov, Writing a PM code,
{\tt  http://astro.uchicago.edu/$\sim$andrey/Talks/PM/pm.pdf}.



\bibitem{2lpt}M. Crocce, S. Pueblas, and R. Scoccimarro, MNRAS \textbf{373},
 369 (2006). 
\bibitem{CMBFAST}U. Seljak, and M. Zaldarriaga,
 Astrophys. J. \textbf{469}, 437 (1996). 

\bibitem{Gadget} V. Springel, MNRAS \textbf{364}, 1105 (2005).


\bibitem{RPTbao}M. Crocce, R. Scoccimarro, Phys. Rev. D  \textbf{77},  023533 (2008). 

\bibitem{HF}R. E. Smith et al., MNRAS \textbf{341}, 1311 (2003)


\bibitem{Khoury}J. Khoury, M. Wyman, arXiv:0903.1292 (2009)


\bibitem{Davis} M. Davis, G. Efstathiou, C. S. Frenk, and
 S. D. M. White, ApJ. \textbf{292}, 371 (1985).  

\bibitem{ShethTormen}R. Sheth, and G. Tormen,  MNRAS \textbf{308}, 119 (1999).

\bibitem{Jenkins} A. Jenkins \textit{et al.}, MNRAS \textbf{321}, 372 (2001). 


\bibitem{PressSchechter} W. Press, and P. Schechter, ApJ \textbf{187}, 425 (1974).


\bibitem{MSS08}M.~C.~Martino, H.~F.~Stabenau, R.~K.~Sheth, Phys. Rev. D  \textbf{79},  084013 (2009).

\bibitem{BBKS}J. Bardeen, J. R. Bond, N. Kaiser, A. Szalay, ApJ. \textbf{304}, 15 (1986).  

\bibitem{CK}S. Cole, N. Kaiser, MNRAS \textbf{237}, 1127 (1989).  

\bibitem{Hui} L. Hui, and K. P. Parfrey, Phys. Rev. D \textbf{77}, 043527 (2008).




\bibitem{SchmidtNB} F. Schmidt,  arXiv:0905.0858  




\bibitem{Feng}X. Feng, private communication. 


\bibitem{Polyanin} A. Polyanin, and V. F. Zaitsev, Ch. 7, \textit{Handbook of
 Nonlinear Partial Differential Equations }, (Chapman \& Hall/CRC, New York, 2004).




\bibitem{NR}W. H. Press, S. A. Teukolsky, W. T. Vetterling, and B. P. Flannery, \textit{Numerical Recipes 3rd Edition: The Art of Scientific Computing}, (Cambridge University Press, Cambridge, 2007).





\bibitem{MathNumericalMA}V. I. Olicker, and L. D. Prussner, Numerische
 Mathematik \textbf{54}, 271 (1988); E. J. Dean, and R. Glowinski, Computer 
Methods in Applied Mechanics and Engineering \textbf{195}, 1344
 (2006). A. M. Oberman, Discrete and Continuous Dynamical Systems Series B
 \textbf{10}, 221 (2008); X. Feng, and M. Neilan, arXiv:0712.1240.   


\bibitem{Shuman}F. G. Shuman, \textit{A Method for Solving the
 Balance Equation}, Technical memorandum No.~6, (Joint Numerical Weather
 Prediction Unit) (1955), {\tt http://docs.lib.noaa.gov/rescue/JNWP/Shuman\_1955 \_May.PDF}.

\bibitem{Meteorologists}  F. G. Shuman, Monthly Weather Review \textbf{85},
 329 (1957). F. H. Bushby, and V. M. Huckle, Quarterly Journal of the Royal
 Meteorological Society \textbf{82}, 409 (1956); G. 'Arnason, Journal of
 Meteorology \textbf{15}, 220 (1957). 








\end{thebibliography}
\end{document}